\pdfoutput=1
\documentclass[journal]{IEEEtran}
\makeatletter
\newcommand\CheckIfSubfigLoaded{\@ifpackageloaded{subfig}{\@latex@warning{The subfig package has been loaded}}{}}
\makeatother

\PassOptionsToPackage{usenames,dvipsnames}{xcolor}

\usepackage{booktabs}
\usepackage{hyperref}
\usepackage[htt]{hyphenat}
\usepackage{xparse}
\ExplSyntaxOn
\NewDocumentCommand{\storedata}{mm}{\tl_set:Nn#1{#2}}
\DeclareExpandableDocumentCommand{\getdata}{O{1}m}{\tl_item:Nn#2{#1}}
\ExplSyntaxOff

\usepackage{filecontents}
\usepackage{datatool}

\usepackage{varwidth} %for the varwidth minipage environment
\usepackage{pdflscape}
\usepackage{fancyhdr}

\renewcommand{\headrulewidth}{0pt}
\renewcommand{\footrulewidth}{0pt}

\fancypagestyle{firstpage}{% Page style for first page
  \fancyhf{}% Clear header/footer
  \renewcommand{\headrulewidth}{0.0pt}% Header rule
  \renewcommand{\footrulewidth}{0.0pt}% Footer rule
	\fancyfoot[C]{UNCLASSIFIED: Distribution Statement A. Approved for public release; distribution is unlimited. \#28420}% Footer

}
\newcommand\Fontvi{\fontsize{6}{7.2}\selectfont}
\usepackage{adjustbox}
\usepackage{array}
\newcolumntype{R}[2]{%
    >{\adjustbox{angle=#1,lap=\width-(#2)}\bgroup}%
    l%
    <{\egroup}%
}
\newcolumntype{L}[2]{%
    >{\adjustbox{angle=#1,lap=\width-(#2)}\bgroup}%
    l%
    <{\egroup}%
    |
}
\newcommand*\rot{\multicolumn{1}{R{70}{1em}}}% no optional argument here, please!
\newcommand*\rotl{\multicolumn{1}{L{70}{2em}}}% no optional argument here, please!

\usepackage{pifont}% http://ctan.org/pkg/pifont
\newcommand{\cmark}{\textcolor{green}{\ding{51}}}%
\newcommand{\xmark}{\textcolor{red}{\ding{55}}}%

\usepackage[autostyle]{csquotes}

\usepackage{amsthm}
\theoremstyle{definition}

\theoremstyle{descrip}
\newtheorem{descrip}{Description:}[section]

\usepackage{amsmath}
\usepackage{cases}
\usepackage{mathrsfs}
\usepackage{amssymb}
\DeclareMathAlphabet\mathbfcal{OMS}{cmsy}{b}{n}
\usepackage{siunitx}

\usepackage{stackengine,graphicx,scalerel}
\newcommand\CircArrowRight[1]{\stackengine{-.3ex}{\scalebox{.8}{#1}}{\CAR}{O}{c}{F}{F}{L}}
\newcommand\CAR{\scaleto{\circlearrowright}{3ex}}

\usepackage[T1]{fontenc}
\usepackage{beramono}
\usepackage{listings}
\lstdefinelanguage{Julia}
  {morekeywords={abstract,break,case,catch,const,continue,do,else,elseif,
      end,export,false,for,function,immutable,import,importall,if,in,
      macro,module,otherwise,quote,return,switch,true,try,type,typealias,
      using,while,define,dynamics,configure,integrate,initOpt,defineTolerances,defineMPC,@NLconstraint,@NLobjective,optimize,allPlots,
      toms,defineIP,simMPC,tomPhase,tomStates,tomControls, icollocate,collocate,integrate,sym2prob,tomRun,initial,final, mpcPlots},
   sensitive=true,
   alsoother={$},
   morecomment=[l]\#,
   morecomment=[n]{\#=}{=\#},
   morestring=[s]{"}{"},
   morestring=[m]{'}{'},
}[keywords,comments,strings]%

\usepackage{tikz,pgfplots,tkz-euclide}

\usetikzlibrary{shapes,arrows.meta,fit,mindmap,trees,calc,quotes,angles}
\usepackage[]{graphicx}

\usepackage{epstopdf}
\usetikzlibrary{matrix}
\usepgfplotslibrary{groupplots,colorbrewer} %http://www.traag.net/2014/06/05/281/
\usetikzlibrary{pgfplots.colorbrewer,datavisualization}

\definecolor{s1}{RGB}{228, 26, 28}
\definecolor{s2}{RGB}{55, 126, 184}
\definecolor{s3}{RGB}{77, 175, 74}
\definecolor{s4}{RGB}{152, 78, 163}
\definecolor{s5}{RGB}{255, 127, 0}

\pgfplotscreateplotcyclelist{my markers}{%
{black, solid, every mark/.append style={solid, fill=black},mark = *},
{s1, dotted, every mark/.append style={solid, fill=s1},mark = halfsquare left*},
{s2, densely dashdotted, every mark/.append style={solid, fill=s2},mark = triangle*},
{s3, loosely dotted, every mark/.append style={solid, fill=s3, scale = 1},mark = diamond*},
{s4, dashed, every mark/.append style={solid, fill=s4, scale = 0.5}, mark = otimes*},  %, mark size = 1
{s5, densely dotted, every mark/.append style={solid, fill=s5},mark = halfcircle*},
{s1,dashdotted, every mark/.append style={solid, fill=s1},mark = 10-pointed star},
{s2,densely dashdotted, every mark/.append style={solid, fill=s2},mark = pentagon},
{s3,solid, every mark/.append style={solid},mark = ball,ball color=s3},
}

\pgfplotscreateplotcyclelist{my markers solid}{%
{black, solid},
{s1, dashed},
{s2, solid},
{s3, solid},
{s4, solid},  %, mark size = 1
{s5, solid},
}

\pgfplotscreateplotcyclelist{my markersB}{%
{s2, densely dashdotted, every mark/.append style={solid, fill=s2},mark = triangle*},
{s3, loosely dotted, every mark/.append style={solid, fill=s3, scale = 1},mark = diamond*},
{s4, dashed, every mark/.append style={solid, fill=s4, scale = 0.5}, mark = otimes*},  %, mark size = 1
{s5, densely dotted, every mark/.append style={solid, fill=s5},mark = halfcircle*},
{s1,dashdotted, every mark/.append style={solid, fill=s1},mark = 10-pointed star},
{s2,densely dashdotted, every mark/.append style={solid, fill=s2},mark = pentagon},
{s3,solid, every mark/.append style={solid},mark = ball,ball color=s3},
}

\pgfplotscreateplotcyclelist{my markersA}{%
{black, solid, every mark/.append style={solid, fill=black},mark = *},
{s1, dotted, every mark/.append style={solid, fill=s1, scale = 0.5},mark = halfsquare left*},
{s2, densely dashdotted, every mark/.append style={solid, fill=s2},mark = triangle*},
{s3, loosely dotted, every mark/.append style={solid, fill=s3, scale = 1},mark = diamond*},
{s4, dashed, every mark/.append style={solid, fill=s4, scale = 0.5}, mark = otimes*},  %, mark size = 1
{s5, densely dotted, every mark/.append style={solid, fill=s5},mark = halfcircle*},
{s1,dashdotted, every mark/.append style={solid, fill=s1},mark = 10-pointed star},
{s2,densely dashdotted, every mark/.append style={solid, fill=s2},mark = pentagon},
{s3,solid, every mark/.append style={solid},mark = ball,ball color=s3},
}

\pgfplotscreateplotcyclelist{my markers ee}{%
{black, solid, every mark/.append style={solid, fill=red},mark = *},
{blue, dotted, every mark/.append style={solid, fill=blue},mark = halfsquare left*},
{densely dashdotted, every mark/.append style={solid, fill=magenta},mark = triangle*},
{loosely dotted, every mark/.append style={solid, fill=violet, scale = 1},mark = diamond*},
{teal, dashed, every mark/.append style={solid, fill=purple}, mark = otimes*},  %, mark size = 1
{red, densely dotted, every mark/.append style={solid, fill=green!60!black},mark = halfcircle*},
{dashdotted, every mark/.append style={solid, fill=green},mark = 10-pointed star},
{densely dashdotted, every mark/.append style={solid, fill=black},mark = pentagon},
{solid, every mark/.append style={solid, fill=red},mark = ball,ball color=green},
}

\pgfplotsset{every axis plot/.append style= ultra thick }
\pgfplotsset{compat=1.5.1}

\pgfplotscreateplotcyclelist{my black white}{%
solid, every mark/.append style={solid, fill=gray}, mark=*\\%
dotted, every mark/.append style={solid, fill=gray}, mark=square*\\%
densely dotted, every mark/.append style={solid, fill=gray}, mark=otimes*\\%
loosely dotted, every mark/.append style={solid, fill=gray}, mark=triangle*\\%
dashed, every mark/.append style={solid, fill=gray},mark=diamond*\\%
loosely dashed, every mark/.append style={solid, fill=gray},mark=*\\%
densely dashed, every mark/.append style={solid, fill=gray},mark=square*\\%
dashdotted, every mark/.append style={solid, fill=gray},mark=otimes*\\%
solid, every mark/.append style={solid},mark=star\\%
densely dashdotted,every mark/.append style={solid, fill=gray},mark=diamond*\\%
}
\usepackage{pgfplotstable}

\pgfplotstableset{
  col sep=comma,
string replace*={_}{\textsubscript},
every head row/.style={before row=\toprule,after row=\midrule},
every last row/.style={after row=\bottomrule},
display columns/0/.style={string type,column name={}},
empty cells with={NaN} % https://tex.stackexchange.com/questions/366739/replace-empty-cells-with-nans-in-pgfplotstable
}

\usepackage{caption,subcaption}
\usepackage{lineno,hyperref}
\modulolinenumbers[5]
\definecolor{links}{HTML}{2A1B81}
\hypersetup{colorlinks,linkcolor=,urlcolor=links}

\tikzstyle{block} = [draw,rectangle,thick,minimum height=2em,minimum width=2em,align=center]
\tikzstyle{sum} = [draw,circle,inner sep=0mm,minimum size=2mm]
\tikzstyle{connector} = [->,thick]
\tikzstyle{output} = [coordinate]

\colorlet{ColorPink}{red!10}
\newcommand{\NLOpt}{\texttt{NLOptControl} }
\newcommand{\NLOptnsp}{\texttt{NLOptControl}}
\newcommand{\NLOptabr}{\texttt{NLOpt}}

\newcommand{\R}{\mathbb{R}}

\newcommand{\TaxName}[1]{\emph{#1}}

\newcommand{\PROPT}{\TaxName{PROPT} }
\newcommand{\PROPTnsp}{\TaxName{PROPT}}
\newcommand{\GPOPSii}{\TaxName{GPOPS-ii} }
\newcommand{\GPOPSiinsp}{\TaxName{GPOPS-ii}}

\newcommand{\DIDO}{\TaxName{DIDO} }

\newcommand{\GPOCS}{\TaxName{GPOCS} }
\newcommand{\GPOCSnsp}{\TaxName{GPOCS}}
\newcommand{\CasADi}{\TaxName{CasADi} }
\newcommand{\CasADinsp}{\TaxName{CasADi}}
\newcommand{\ACADO}{\TaxName{ACADO} }
\newcommand{\ACADOnsp}{\TaxName{ACADO}}
\newcommand{\JuMP}{\TaxName{JuMP} }
\newcommand{\JuMPnsp}{\TaxName{JuMP}}
\newcommand{\Julia}{\TaxName{Julia} }
\newcommand{\Juliansp}{\TaxName{Julia}}

\newcommand{\Fortran}{\TaxName{Fortran} }
\newcommand{\Fortrannsp}{\TaxName{Fortran}}
\newcommand{\C}{\TaxName{C} }
\newcommand{\Cnsp}{\TaxName{C}}
\newcommand{\Cpp}{\TaxName{C++} }
\newcommand{\Cppnsp}{\TaxName{C++}}
\newcommand{\MATLAB}{\TaxName{MATLAB} }
\newcommand{\MATLABnsp}{\TaxName{MATLAB}}

\newcommand{\IPOPT}{\TaxName{IPOPT} }
\newcommand{\IPOPTnsp}{\TaxName{IPOPT}}
\newcommand{\KNITRO}{\TaxName{KNITRO} }
\newcommand{\KNITROnsp}{\TaxName{KNITRO}}
\newcommand{\SNOPT}{\TaxName{SNOPT} }

\newcommand{\AMPL}{\TaxName{AMPL} }
\newcommand{\GAMS}{\TaxName{GAMS} }
\newcommand{\ReverseDiffSparse}{\TaxName{ReverseDiff} }
\newcommand{\ReverseDiffSparsensp}{\TaxName{ReverseDiff}}

\newcommand{\DifferentialEquations}{\TaxName{DifferentialEquations} }

\newcommand{\matplotlib}{\TaxName{matplotlib} }
\newcommand{\GR}{\TaxName{GR} }

\newcommand{\Huck}[1]{\if\showComments1\textcolor{blue}{[#1]} \else{}\fi}
\newcommand{\Stein}[1]{\if\showComments1\textcolor{red}{[#1]} \else{}\fi}
\newcommand{\Tulga}[1]{\if\showComments1\textcolor{teal}{[#1]} \else{}\fi}
\newcommand{\benoit}[1]{\if\showComments1\textcolor{orange}{[#1]} \else{}\fi}

\def\showComments{0}
\def\slides{0}
\def\showoutline{0}
\def\doublecol{1}
\def\thesis{0}

\newcommand{\nla}{\NLOptabr$_{\text{LGR}_1}$ }
\newcommand{\nlansp}{\NLOptabr$_{\text{LGR}_2}$}
\newcommand{\nlb}{\NLOptabr$_{\text{LGR}_2}$ }
\newcommand{\nlbnsp}{\NLOptabr$_{\text{LGR}_2}$}
\newcommand{\nlc}{\NLOptabr$_{\text{LGR}_4}$ }
\newcommand{\nlcnsp}{\NLOptabr$_{\text{LGR}_4}$}
\newcommand{\nleu}{\NLOptabr$_\text{E}$ }
\newcommand{\nleunsp}{\NLOptabr$_\text{E}$}
\newcommand{\nltr}{\NLOptabr$_\text{T}$ }
\newcommand{\nltrnsp}{\NLOptabr$_\text{T}$}

\newcommand{\propta}{\PROPTnsp$_{\text{C}_1}$ }
\newcommand{\proptansp}{\PROPTnsp$_{\text{C}_1}$}
\newcommand{\proptb}{\PROPTnsp$_{\text{C}_2}$ }
\newcommand{\proptbnsp}{\PROPTnsp$_{\text{C}_2}$}
\newcommand{\proptc}{\PROPTnsp$_{\text{C}_4}$ }
\newcommand{\proptcnsp}{\PROPTnsp$_{\text{C}_4}$}

\storedata\NLstringsA{{solution}{coll. pts.}}
\storedata\NLstringsC{{\NLOpt}{\texttt{NLOpt.} colloc. pts.}}
\storedata\PRstringsA{{\PROPT}{\PROPT colloc. pts.}}
\storedata\NLstringsB{{\nla}{\nlb}{\nlc}{\nltr}{\nleu}}
\storedata\PRstringsB{{\propta}{\proptb}{\proptc}}
\storedata\analytic{{Analytic solution}}

\storedata\MPCstringsA{{Case A}{Case B}{Case C}{Case D}}
\storedata\MPCstringsB{{Plant}}

\if\slides0
  \newcommand{\ha}{3.3cm}
  \newcommand{\wa}{6cm}
  \newcommand{\hb}{3.7cm}
  \newcommand{\wb}{6cm}

\fi
\usepackage{outlines}
\usepackage{enumitem}
\setenumerate[1]{label=\Roman*.}
\setenumerate[2]{label=\Alph*.}
\setenumerate[3]{label=\roman*.}
\setenumerate[4]{label=\alph*.}

\begin{document}

%\thanks{Manuscript submitted October, 2018. The authors wish to acknowledge the financial support of the Automotive Research Center (ARC) in accordance with Cooperative Agreement W56HZV-14-2-0001 U.S. Army Tank Automotive Research, Development and Engineering Center (TARDEC) Warren, MI.,DISTRIBUTION A. Approved for public release; distribution unlimited.}%
%\thanks{H. Febbo, J. L. Stein, and T. Ersal are with the Department of Mechanical Engineering, University of Michigan, Ann Arbor, MI 48109 USA (e-mail: febbo@umich.edu; stein@umich.edu; tersal@umich.edu).}%
%\thanks{P. Jayakumar is with the U.S. Army RDECOM-TARDEC, Warren, MI 48397 (email: paramsothy.jayakumar.civ@mail.mil)}%
%\thanks{$^{*}$Corresponding author (tersal@umich.edu)}}

%\begin{frontmatter}
%https://www.elsevier.com/__data/assets/pdf_file/0008/56843/elsdoc-1.pdf
\title{\NLOptnsp: A modeling language for solving optimal control problems}

\fancyhf{}
\fancyfoot[C]{The authors wish to acknowledge the financial support of the Automotive Research Center (ARC) in accordance with Cooperative Agreement W56HZV-14-2-0001 U.S. Army Tank Automotive Research, Development and Engineering Center (TARDEC) Warren, MI., UNCLASSIFIED: Distribution Statement A. Approved for public release; distribution is unlimited. \#28420}% Footer
\renewcommand{\headrulewidth}{0pt}
\renewcommand{\footrulewidth}{0pt}

\fancypagestyle{firstpage}{% Page style for first page
  \fancyhf{}% Clear header/footer
  \renewcommand{\headrulewidth}{0.0pt}% Header rule
  \renewcommand{\footrulewidth}{0.0pt}% Footer rule
	\fancyfoot[C]{UNCLASSIFIED: Distribution Statement A. Approved for public release; distribution is unlimited. \#28420}% Footer

}

%\tnotetext[t2]{The second title footnote which is a longer
%text matter to fill through the whole text width and
%overflow into another line in the footnotes area of the
%first page.}
%\author{Huckleberry~Febbo,Paramsothy~Jayakumar,Jeffrey~L.~Stein,Tulga~Ersal\corref{cor1}\ %}
\author{Huckleberry~Febbo, Paramsothy~Jayakumar, Jeffrey~L.~Stein, and Tulga~Ersal$^{*}$%
\thanks{The authors wish to acknowledge the financial support of the Automotive Research Center (ARC) in accordance with Cooperative Agreement W56HZV-14-2-0001 U.S. Army Tank Automotive Research, Development and Engineering Center (TARDEC) Warren, MI.,DISTRIBUTION A. Approved for public release; distribution unlimited.}%
\thanks{H. Febbo, J. L. Stein, and T. Ersal are with the Department of Mechanical Engineering, University of Michigan, Ann Arbor, MI 48109 USA (e-mail: febbo@umich.edu; stein@umich.edu; tersal@umich.edu).}%
\thanks{P. Jayakumar is with the U.S. Army RDECOM-TARDEC, Warren, MI 48397 (email: paramsothy.jayakumar.civ@mail.mil)}%
\thanks{$^{*}$Corresponding author (tersal@umich.edu)}}
\maketitle

\if\thesis0
\begin{abstract}
\fi
Current direct-collocation-based optimal control software is either easy to use or fast, but not both.
This is a major limitation for users that are trying to formulate complex optimal control problems (OCPs) for use in on-line applications.
This paper introduces \NLOptnsp, an open-source modeling language that allows users to both easily formulate and quickly solve nonlinear OCPs using direct-collocation methods.
To achieve these attributes, \NLOpt (1) is written in an efficient, dynamically-typed computing language called \Juliansp, (2) extends an optimization modeling language called \JuMP to provide a natural algebraic syntax for modeling nonlinear OCPs; and (3) uses reverse automatic differentiation with the acyclic-coloring method to exploit sparsity in the Hessian matrix.
This work explores the novel design features of \NLOpt and compares its syntax and speed to those of \PROPTnsp.
The syntax comparisons shows that \NLOpt models OCPs more concisely than \PROPTnsp.
%  TODO Elaborate on Bolza formds
The speeds of various collocation methods within \PROPT and \NLOpt are benchmarked over a range of collocation points using performance profiles; overall, \NLOptnsp's single, two, and four interval pseudospectral methods are roughly $14$, $26$, and $36$ times faster than \PROPTnsp's, respectively.
%
%These profiles show that a Legendre-Gauss-Radau quadrature collocation method with one, two, and four intervals implemented in \NLOpt solves an obstacle avoidance problem for a ground vehicle $200$, $400$ and $500$ times faster than Chebyshev quadrature collocation method with one, two, and four phases implemented in \PROPTnsp, respectively.
%
%
%The approach taken to further specialize \JuMP to include syntax for modeling OCPs holds great potential for other classes of optimization problems.
%Further specializing \JuMP to include syntax for easily modeling and quickly solving OCPs
%Overall, \NLOpt is
%results indicate that performance profiles can help select an appropriate collocation method for a particular type of problem and that \NLOpt is faster than \PROPTnsp.
%
\NLOpt is well-suited to improve existing off-line and on-line control systems and to engender new ones.

\if\thesis0
\end{abstract}
\fi

% \if\thesis0
% \begin{keyword}
% %Model Predictive Control \sep Optimal Control \sep Gaussian Quadrature \sep Automatic Differentiation \sep Nonlinear Programming Problem \sep Direct Methods \sep acyclic-coloring \sep \NLOpt
% optimal control \sep nonlinear model predictive control \sep modeling language \sep direct collocation \sep acyclic-coloring \sep reverse automatic differentiation \sep performance profiles \sep \NLOpt
% \end{keyword}
% \fi

\section{Introduction}
Optimal control software packages that implement direct-collocation methods are used in a number of off-line \cite{patterson2014gpops,gpops_spaceA,Rutquist2016,jorris2008constrained,kang2007pseudospectral,casadi_chemicalA} and on-line \cite{casadi_robotA,frasch2013auto} applications as summarized in Table \ref{tab:solvers}.
The primary function of these packages is to directly transcribe a human modeler's formulation of an optimal control problem (OCP) into a nonlinear programming problem (NLP). A key challenge with this process is enabling human modelers (i.e., users) to easily formulate new and complex problems while producing an NLP that can be quickly solved by an external NLP solver. However, current direct-collocation-based optimal control software packages are generally either fast or easy to use, but not both. Thus, these package are not well suited for non-expert users trying to formulate complex problems for on-line applications, wherein speed is critical. Therefore, there is a need for a direct-collocation-based optimal control software package that is both fast and easy to use. In this paper, an approach to bridging this gap is presented and incorporated into a new, open-source optimal control modeling language called \NLOpt \cite{febbo_2017}.

%
%As described in Section~\ref{sec:approach}, if package has an algebraic syntax that closely resembles the Bolza form of OCPs \cite{kelly2017introduction} it is categorized as easy to use.
%  TODO Elaborate on Bolza form
% modeling errors... what else... overly flexible.., but we still consider many as easy to use...
% consider elaborating on moving section
%%%%%%%%%%%%%%%%
%
As seen in Table \ref{tab:solvers}, some of the most well-known optimal control software packages (\GPOPSii , \PROPT, \DIDO) are closed-source and often require a licensing-fee. These drawbacks limit their research value, since they are not freely available to the entire research community, results may be difficult to reproduce, and if the details of the underlying algorithms cannot both be seen and modified, then open validation and development of the these algorithms is not possible \cite{rao2010algorithm,becerra2010solving}. Fortunately, several noteworthy open-source optimal control software packages exist. For completeness, this paper does not limit its discussions to these open-source packages.

Optimal control packages with an algebraic syntax that closely resembles the Bolza form of OCPs \cite{kelly2017introduction} are categorized as easy to use.
It is noted that there are other design features that affect ease of use; for instance, not having a built-in initialization algorithm \cite{betts2003initialization} reduces ease of use, but these aspects of ease of use are not addressed in this paper.
Table \ref{tab:solvers} shows that this work categorizes the direct-collocation-based optimal control software packages \GPOPSii \cite{patterson2014gpops}, \PROPT \cite{Rutquist2016} and \GPOCS \cite{rao2007user} as easy to use and \CasADi \cite{andersson2012casadi} and \DIDO \cite{ross2007beginner} as not easy to use.

\if\doublecol1
\newcommand{\litW}{0.48}
\else
\newcommand{\litW}{0.9}
\fi
\begin{table}
 \begin{center}
 \resizebox{\litW\textwidth}{!}{%
 \begin{tabular}{l|ccc|ccc|ccc}
 & \multicolumn{6}{c|}{Applications}
 & & & \\
 \cline{2-10}
 & \multicolumn{3}{c|}{Off-line} & \multicolumn{3}{c|}{\textcolor{black}{On-line}}
 & \multicolumn{3}{c}{ Properties}  \\
 \cline{2-10}
 Software &
 \rot{Chemical} & \rot{Space vehicle} & \rotl{Medical} &
 \rot{Air vehicle} & \rot{Ground vehicle} & \rotl{Robot} &
 \rot{Open-source}  & \rot{Easy to use} & \rot{\textcolor{black}{Fast}} \\
 \hline
 \GPOPSii &
 \cite{patterson2014gpops}&  \cite{patterson2014gpops,gpops_spaceA}  &  &
  &  & \textcolor{black}{\cite{gpops_robotA}}\textcolor{red}{$\dagger$} &
  \xmark  &\cmark   &\xmark \\
 \PROPT  &
 \cite{Rutquist2016} & \cite{Rutquist2016}  & \cite{Rutquist2016}  &
  &  & &
  \xmark &\cmark   &\xmark  \\
 \GPOCS &
 &\cite{jorris2008constrained} & &
 \textcolor{black}{\cite{basset2010computing}}\textcolor{red}{$\dagger$} & & &
 \cmark  & \cmark &\xmark  \\
 \DIDO &
 & \cite{kang2007pseudospectral}& &
 \textcolor{black}{\cite{basset2010computing}}\textcolor{red}{$\dagger$} & & &
 \xmark & \xmark&\xmark  \\
 \ACADO &
  & & &
  & \cite{frasch2013auto} & &
  \cmark  &\xmark  &\cmark   \\
 \CasADi &
  \cite{casadi_chemicalA}& & &
  &  & \cite{casadi_robotA} &
  \cmark &\xmark  &\cmark   \\
 \textcolor{black}{Custom} &
  & & &
 & \textcolor{black}{\cite{liu2017combined,febbo2017moving}}\textcolor{red}{$\dagger$} & &
 \xmark  & \xmark& \xmark\\
 \NLOpt &
  & & &
  & \cite{febbo_2017} & &
  \cmark  & \cmark& \cmark\\
 \hline
 \end{tabular}
 }
 \end{center}
\caption{Landscape of direct-collocation-based optimal control software focusing on their applications and properties. \textcolor{red}{$\dagger$} indicates that the software is too slow for use the on-line application.} \label{tab:solvers}
\end{table}

%%%%%%%%%%%%
For ease of use, modeling languages should have a syntax that closely resembles the class of problems for which they have been designed.
Modeling languages like AMPL and GAMs are not embedded in a pre-existing computational language, which allows for syntactical flexibility, when developing them.
However, this approach (1) makes development of the modeling language difficult and time-consuming, and (2) does not directly expose users to the breath of features available in a computational language such as \Cpp or \MATLABnsp.
For these reasons, modeling languages are often embedded in a pre-existing computational language.

%%%%%%%%%%%%%%%
%Embedding a modeling language in a pre-existing language brings its own issue.
It can be difficult to establish a syntax for the modeling within the syntactical confines of a pre-existing computational language.
To overcome this issue, operator overloading can be used.
For instance, a multiple-shooting method based optimal control software package called \ACADO \cite{houska2011acado} uses operator overloading to allow its user to define an OCP using symbolic expressions that closely resemble the actual mathematical expressions of the problem.
However, a naive implementation of operator overloading can lead to performance issues \cite{LubinDunningIJOC}.
%
%Additionally, Moritz Dielhl, a researcher who developed \ACADO and MUSCOD-II, later acknowledges that \blockquote{...\ACADO Toolkit \cite{houska2011acado}, DIRCOL \cite{von2001dircol}, DyOS \cite{dyos}, and MUSCOD-II \cite{diehl2001muscod} --- the usage of the of these tools, in particular for a user not directly involved with their actual development, is typically limited to comparably restrictive problem formulations. \cite{andersson2012casadi}}.
%\Tulga{of the of these .. Does this typo exist in the original language? If so, add [sic] after the typo. }
%\Stein{paraphrase this.}
Additionally, Moritz Dielhl, a researcher who developed \ACADO and MUSCOD-II, later acknowledges that, \ACADO Toolkit \cite{houska2011acado}, DIRCOL \cite{von2001dircol}, DyOS \cite{dyos}, and MUSCOD-II \cite{diehl2001muscod} restrict the problem formulations, particularly for users not involved with the development of these tools \cite{andersson2012casadi}.
%%%%%%%%%%
%The above quote is from a paper \cite{andersson2012casadi} that introduces \CasADinsp.
The above acknowledgment is included in a paper \cite{andersson2012casadi} that introduces \CasADinsp.
\CasADi allows users to formulate OCPs with fewer restrictions that \ACADOnsp.
However, \CasADi requires that users write the code for the transcription methods.
Transcription methods are a general class of numerical methods used to approximate continuous-time OCPs; a direct-collocation method is a type of transcription method.
\CasADi lets users to code their own transcription methods to avoid creating a "black box" OCP solver that is only capable of solving restrictive formulations, as with \ACADOnsp.
While this approach may be pedagogically valuable for users, it can lead to bugs and long development time \cite{leek2016optimal} and it makes \CasADinsp's syntax not closely resemble OCPs.
For these reasons, this paper does not categorize \CasADi as easy to use.
On similar grounds, \DIDO is not categorized as easy to use.

%%%%%%%%%%%%%%%%%%%%%%
For safety in on-line applications, the trajectory needs to be provided to the plant in real-time.
An on-line optimal control example is a nonlinear model predictive control (NMPC) problem.
Real-time is achieved when the NLP solve-times are all less than the chosen execution horizon.
Otherwise, the low-level controllers will not have a trajectory to follow.
Despite the need for small solve-times (i.e., speed), several implementations of direct-collocation methods within the \MATLAB computational language are not able to achieve solve-times that are less than the execution horizon for a number of NMPC applications.
As seen in Table \ref{tab:solvers}, \GPOCSnsp, \GPOPSiinsp, and custom \MATLAB software are not fast enough for NMPC applications in aircraft \cite{basset2010computing}, robot \cite{gpops_robotA}, and UGV \cite{liu2017combined,febbo2017moving} systems, respectively.
On the other hand, \CasADinsp, which is written in \Cppnsp, is fast enough for an NMPC application in a robot system \cite{casadi_robotA}.
Given this practical limitation, this paper will now discuss why some direct-collocation-based optimal control packages are fast while others are slow.

%%%%%%%%%
As seen in Table \ref{tab:solvers}, this work categorizes \GPOPSiinsp, \PROPTnsp, \GPOCSnsp, and \DIDO as slow and \CasADi as fast.
\benoit{This sentence seems weird. It has already been said and its purpose is not clear as the beginning of this paragraph}
If a package uses sparse automatic differentiation methods implemented in a computation language that approaches the speeds of \Cnsp, it is categorized as fast; the reasoning for this categorization is explained below.

%%%%%%%%%%%%%%%%%%%%%%%%%%%%%%%%%%%%%%%%%%%%%%%%%%%%
The main algorithmic step in direct method based numerical optimal control is solving the NLP.
The solve-time for this step consists of two major parts: (1) the time spent running optimization algorithms within the NLP solver, and (2) the time spent evaluating the nonlinear functions and their corresponding derivatives.
Fortunately, low-level algorithms, which are available within several prominent NLP solvers, such as \KNITRO \cite{Byrd2006}, \IPOPT \cite{AndreasWaechter2004}, and \SNOPT \cite{Gill2002}, can be used to reduce the time associated with running the optimization algorithms.
%The second component of the solve-time can be reduced by using a fast sparse differentiation method implemented in a fast computational language.
The second component is discussed here in terms of current direct-collocation-based optimal control software packages.

%%%%%%%%%%%%%%%%%%%%%%%%%%%%%%%%%%%%%%%%%%%%%%%%%%%%
The speed of direct-method-based optimal control software depends on the speed of the differentiation method within the computational language in which it is implemented.
\GPOPSii uses a sparse finite difference method \cite{patterson2012exploiting} to calculate the derivatives using the \MATLAB computational language.
However, finite difference methods, like the sparse finite difference method, are not only slow, but they are also inaccurate \cite{tenny2002feasible}.
%%% TODO add citation above
In addition to this, the dynamically-typed \MATLAB computational language is typically slow in comparison to statically-compiled languages such as \C and \Fortrannsp.
Since \GPOPSii uses a slow differentiation method within a relatively slow computational language, it is categorized as slow.
\PROPT uses either symbolic- or forward-automatic differentiation to calculate the derivatives using \MATLABnsp.
While \PROPTnsp's methods are more accurate and generally faster than finite difference methods, they do not exploit the sparse structure of the Hessian matrices that is born from a direct-collocation method, like the sparse finite difference method in \GPOPSiinsp.
Given this computational limitation and the slow speed of \MATLABnsp, this paper considers \PROPT to be slow as well.
On the other hand, \CasADi uses the star-coloring method \cite{gebremedhin2005color} to exploit the sparse structure of the Hessian matrix and reverse automatic differentiation implemented in \Cpp \cite{andersson2012casadi}.
Since \CasADi employs a differentiation methods that is well suited for the sparse structure of the Hessian matrix and it is implemented in a fast computational language, \CasADi is categorized as fast.
On similar grounds, this paper identifies \GPOCS and \DIDO as slow.

In sum, there is no direct-collocation-based optimal control software package is both fast and easy to use.
%
%The packages that are considered to be easy to use (i.e., \GPOPSiinsp, \PROPTnsp, and \GPOCSnsp) are not fast, and the packages that are considered to be fast (i.e., \CasADinsp) are not easy to use.
\CasADi is fast, but not easy to use; and \GPOPSiinsp, \PROPTnsp, and \GPOCS  are easy to use, but not fast.
Thus, there is a need for a package that is both fast and easy to use.

This paper investigates an approach for improving both speed and ease of use of optimal control software.
%
%This approach is described in detail in Section~\ref{sec:approach}. Here it is noted that this approach uses recent advances in computational languages and differentiation methods.
As described in detail in Section~\ref{sec:approach}, this approach uses recent advances in computational languages and differentiation methods in contrast to the computational languages and differentiation methods used by current direct-collocation-based optimal control software. Additionally, also unlike current direct-collocation-based optimal control software packages, this approach extends an optimization modeling language to include syntax for modeling OCPs. More specifically, this approach is as follows:
\begin{center}
\textbf{Approach}
\end{center}
\begin{itemize}
 \item For ease of use and speed, \NLOpt is embedded in the fast, dynamically-typed \Julia programming language \cite{bezanson2012julia}.
 \item For increased ease of use, \NLOpt extends the \JuMP optimization modeling language \cite{DunningHuchetteLubin2017}, which is written in \Juliansp, to include a natural syntax for modeling OCPs in Bolza form.
 \item For increased speed, \NLOpt uses the acyclic-coloring method \cite{griewank1996algorithm} to exploit sparsity in the Hessian matrix and reverse-automatic differentiation through the \ReverseDiffSparse package \cite{revels_2017}, which is also written in \Juliansp.
\end{itemize}
Therefore, this work addresses the following research question: Can the above outlined approach improve speed and ease of use of direct-collocation-based optimal control software?
% TODO: consider adding: , without sacrificing reliability?
%
This question is answered by comparing \NLOptnsp's speed and ease of use to those of \PROPTnsp.

\NLOpt was released as a free, open-source software package in the summer of $2017$ \cite{febbo_2017}.
Since then, the literature has shown that \NLOpt is fast and easy to use.
For speed, \NLOpt was leveraged to solve complex trajectory planning problems for an unmanned ground vehicle system in real-time --- solving these types of problems in real-time using \MATLAB was not feasible in prior work \cite{liu2017combined,febbo2017moving}.
For ease of use, \NLOpt was used to create a new optimal control based learning algorithm \cite{lessard2018optimal} without any help from the developers of \NLOptnsp.

The remainder of this paper is organized as follows.
Section~\ref{sec:approach} further describes \NLOptnsp's approach to bridging the research gap.
Section~\ref{sec:scope} describes the classes of off-line and on-line OCPs that can be solved using \NLOptnsp.
Section~\ref{sec:direct} provides a brief background on numerical optimal control and a mathematical description of the direct-collocation methods implemented within \NLOptnsp.
%
%Section~\ref{sec:design} details the organization of \NLOptnsp's data, its user interface, its visualization capabilities, and finally, its external software libraries are formally acknowledged.
%
Section~\ref{sec:results} provides an example that compares \NLOptnsp's ease of use against \PROPTnsp's and benchmarks \NLOptnsp's speed against \PROPTnsp.
Section~\ref{sec:discussion} answers the research question and discusses further implications.
Finally, Section~\ref{sec:conclusion} summarizes the work and draws conclusions.
\section{Software ecosystem}\label{sec:approach}
%Software advances in computational languages, optimization modeling languages, and differentiation methods and tools are the foundational building blocks coupled with the author's insights into numerical optimal control and the limitations of the current software in the this area, particularly related to on-line ground vehicle applications led to the development of \NLOptnsp.
Advances in computational languages, optimization modeling languages, and differentiation methods and tools made it possible to create \NLOptnsp.
This section describes these software advances and shows how they can be leveraged to create a modeling language for a class of optimization problems.
%extend the \JuMP optimization modeling language to include new classes of optimization problems.
% NOT talking about julia as an advance

\subsection{Computational languages}
Direct-collocation based optimal control software packages are embedded in either a statically- or a dynamically typed computational language.
%Dynamically typed computational languages are typically considered to be slow, yet easy to use, while statically typed languages are considered to be fast, and yet hard to use.
Dynamically typed languages enable users to quickly develop and explore new concepts, yet they are typically slow; statically typed languages sacrifice the user's productivity for speed.
Recently, however, a dynamically typed computing language called \Julia has become a popular alternative to the computing languages that the current optimal control software packages are embedded in.
It has become popular, because it allows users to write high-level code that closely resembles their mathematical formulas, while producing low-level machine code that approaches the speed of \C and is often faster than \Fortran \cite{bezanson2012julia}.
\benoit{With approaches, the reader may think there is a gap while there should not. Julia code is compile just like C code, the assembly produced might just be the same (it is even more true if Clang is used instead of GCC to compile C) so there should be no gap.}
The claim that \Julia is not only fast, but also easy to use, motivates the investigation presented in this paper.
Specifically, this paper investigates the ability of the \Julia computational language to improve speed and ease of use for optimal control software.

\subsection{Modeling optimization problems}
%M:
In the late 1970's, researchers using optimization software were more concerned with the need to improve the software's ease of use than its speed \cite{fourer2012evolution}.
Eventually, this concern led to the development a number of optimization modeling languages, such as \GAMS \cite{rosenthal2004gams} and \AMPL \cite{fourerampl}.
\benoit{I would consider GAMS and AMPL as both easy to use and fast (see https://epubs.siam.org/doi/pdf/10.1137/15M1020575). The fact that it is not written in an existing language makes it more difficult to embed it into a larger project or to write extension to it (which is why JuMP is better ^^).
In this sentence, it seems that you are categorizing GAMS, AMPL as slow.}
The role of an optimization modeling language is to translate optimization problems from a human-friendly language to a solver-friendly language \cite{udell2014convex,fragniere2000structure}.
In other words, optimization modeling languages do not solve optimization problems; they focus on modeling problems at a high-level and passing optimization problems to external low-level solvers, which are the NLP solvers and the differentiation tools in the context of this work.
Similarly, in this work, the high-level problem is the NLP, given in Eqn. \ref{eq:cost_nlp} - Eqn. \ref{eq:eq_nlp} (i.e., the NLP model) as
\begin{align}
   & \underset{ z \in \R^n}{\text{minimize}}
   && f(z) \label{eq:cost_nlp} \\
   & \text{subject to}
   && g(z) \leq 0  \label{eq:ineq_nlp}\\
   &&& h(z) = 0 \label{eq:eq_nlp}
\end{align}
where the objective function $f: \R^n \rightarrow \R$, with $n$ defined as the number of design variables; the inequality constraints $g: \R^n \rightarrow \R^e$; and the equality constraints $h: \R^n \rightarrow \R^q$, are all assumed to be twice-continuously differentiable functions \cite{AndreasWaechter2004, Betts2010}.
%The low-level solvers are the NLP solvers and the differentiation tool.
%

A number of standard optimization problem classes do not fit readily into the NLP model.
In addition to this, translating these standard problem classes into the NLP model can require significant work.
Thus, for users interested in simply modeling these standard problem classes, and not translating these problems into an NLP model, the NLP model should be extended to include higher-level modeling languages for these standard problem classes.
However, most optimization modeling languages are not designed to be extended in this fashion \cite{fragniere2000structure}.
Because of this limitation, both the speed and ease of use of optimal control packages have suffered.
%
%So, to solve these standard optimization problems, software packages are often developed to interface directly with NLP solvers with inefficient differentiation methods (\GPOCSnsp, \GPOPSiinsp, and \PROPTnsp).
%the state-of-the-art differentiation methods that are available through optimization modeling languages are not available in
\GPOCSnsp, \GPOPSiinsp, and \PROPT are slow because the sparse-automatic differentiation methods --- typically available through an optimization modeling language --- are not available in \MATLABnsp; so, these packages use less efficient differentiation methods.
Additionally, since these packages are not built upon an existing NLP modeling language, the API tends to be overly flexible, which can lead to modeling errors \cite{houska2011acado}.

\JuMP \cite{LubinDunningIJOC}, a recent optimization modeling language that is embedded in the fast, dynamically-typed \Julia programming language \cite{bezanson2012julia}, is designed to be extended to include new classes of optimization problems.
\JuMP extensions include: parallel multistage stochastic programming \cite{huchette2014parallel}, robust optimization \cite{dunning2016advances}, chance constraints \cite{lubin2016robust}, and sum of squares \cite{benoit_2019}.
Moreover, \JuMP provides an interface for both the \KNITRO and \IPOPT NLP solvers as well the \ReverseDiffSparse differentiation tool.
\ReverseDiffSparse \cite{revels_2017} is also embedded in the \Julia programming language and utilizes reverse automatic differentiation with the acyclic-coloring method \cite{gebremedhin2009efficient} to exploit sparsity in the Hessian matrices.
Research shows that the acyclic-coloring method is faster than the star-coloring method \cite{gebremedhin2009efficient}, which was used in the \CasADi package \cite{andersson2012casadi}.
These advances are leveraged to create an optimal control modeling language called \NLOptnsp.

\subsection{Proposed software ecosystem}
\begin{figure}
\begin{minipage}[c]{\linewidth}
 \begin{center}
  \includegraphics[width=0.75\textwidth]{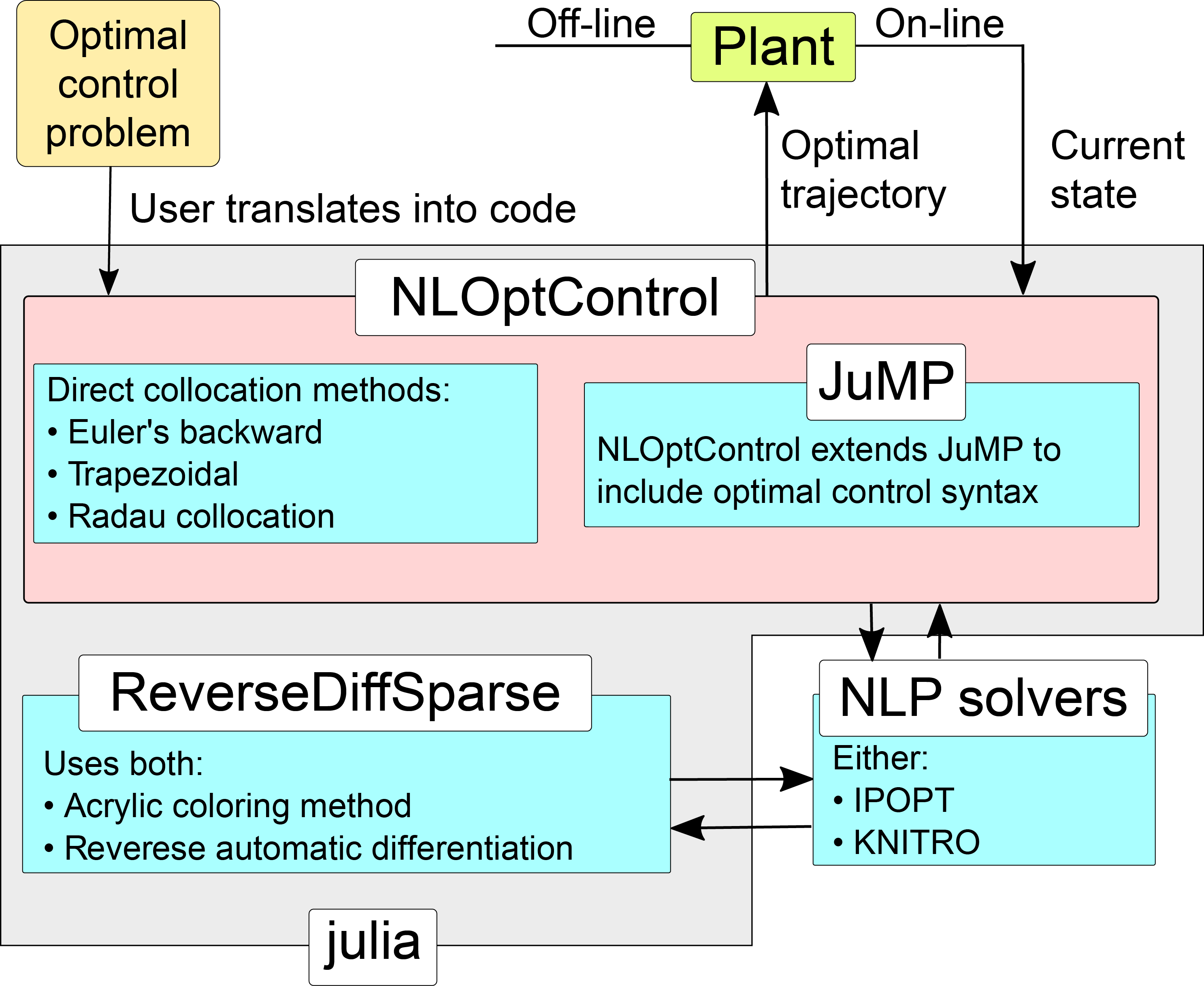}  \caption{Proposed software framework for nonlinear OCPs.} \label{fig:framework} %
  \end{center}
\end{minipage}
\end{figure}

Fig. \ref{fig:framework} presents \NLOptnsp's software ecosystem and its function as an optimal control software package.
In terms of this ecosystem, it is: embedded in \Juliansp; extends \JuMP to provide a natural syntax for modeling OCPs; leverages \ReverseDiffSparsensp; and interfaces with \KNITROnsp, \IPOPTnsp, and potentially other solvers to solve the automatically formulated NLP problem.
To use \NLOptnsp, users need only formulate their OCP into a syntax-based model of the OCP.
This model is then approximated using one of the direct-collocation methods implemented in \NLOptnsp, which at the time of this writing include: the Euler's backwards, the trapezoidal, and the Radau collocation methods.
After the model has been approximated, the software ecosystem solves this approximation to determine an optimal trajectory.
This trajectory can then be followed using low-level controllers to control the plant for either an off-line or on-line tasks.
\section{Scope of \NLOpt}\label{sec:scope}
\NLOpt is designed for modeling OCPs and solving them for either off-line or on-line applications.
%
%At the heart of an NMPC problem lies a OCP.
This section shows the types of problems that \NLOpt can model, and demonstrates \NLOptnsp's visualization capabilities and salient design features for on-line applications (e.g., NMPC problems).
%Afterwards, \NLOptnsp's functionality tailored for on-line applications (i.e., NMPC problems)

\subsection{Modeling OCPs}
An important class of optimization problems is the OCP.
\NLOpt models single-phase, continuous-time, OCP in a Bolza form  \cite{kelly2017introduction} that is tailored for NMPC problems and adds slack constraints on the initial and terminal states as
\if\slides1
\Fontvi

\begin{columns}

%\onslide<1->{
\begin{column}{0.45\textwidth}
\begin{equation*}
  \begin{aligned}
    & \underset{x(t),u(t),{x_0}_s,{x_f}_s\;t_f}{\textbf{minimize}} & \\
    \hline
    %\onslide<3->{
    &{\cal{M}}(x(t_0+t_{ex}),t_0+t_{ex},x(t_f),t_f) & \textbf{\color{blue}Mayer term} \\
    & + \int_{t_0+t_{ex}}^{t_f}L(x(t),u(t),t) \, \text{d}t  &\textbf{\color{blue}Lagrangian term}\\
    & +  w_{s0}{x_0}_s + w_{sf}{x_f}_s & \textbf{\color{blue}Slack term}\\
    %\onslide<4->{
     &\text{subject to} & \\
    \hline
    &  \frac{\text{d}x}{\text{d}t}(t) - F(x(t),u(t),t) =0  & \textbf{\color{blue}Dynamics}\\
    & C(x(t),u(t),t) \le 0  & \textbf{\color{blue}Path} \\
    &    x_{min} \leq  x(t) \leq  x_{max}  &  \textbf{\color{blue}State limits}\\
    &  u_{min} \leq  u(t) \leq  u_{max} &  \textbf{\color{blue}Control limits}\\
    &   t_{f_{min}} \leq  t_f \leq  t_{f_{max}}  & \textbf{\color{blue}Final time}\\
    &  x_0-x_{0_{tol}} \leq x(t_0+t_{ex}) \leq x_0+x_{0_{tol}} &  \textbf{\color{blue}Initial state}\\
    & x_0 - x(t_0+t_{ex}) \leq {x_0}_s  & \textbf{\color{blue}Initial state slack}\\
    & x_0 + x(t_0+t_{ex}) \geq  {x_0}_s   & \textbf{\color{blue}}\\
    &  x_f-x_{f_{tol}} \leq x(t_f) \leq x_f+x_{f_{tol}}  & \textbf{\color{blue}Terminal state}\\
    & x_f - x(t_f) \leq {x_f}_s & \textbf{\color{blue}Terminal state slack}\\
    & x_f + x(t_f) \geq  {x_f}_s  & \textbf{\color{blue}}
  \end{aligned}
\end{equation*}
\end{column}
\hspace{0.5cm}
%}}}

%\onslide<2->{
\begin{column}{0.45\textwidth}
where,
\Fontvi
\begin{center}
\begin{tabular}{c l}
\hline
$t$ & time \\
$t_f$ & variable final time \\
$x(t) \in \R^{n_{st}}$ & state \\
$u(t) \in \R^{n_{ctr}}$ &  control \\
$n_{st}$ & number of states \\
$n_{ctr}$ & number of controls \\
${x_0}_s$ and ${x_f}_s$ & initial and terminal slack variables \\
$t_0$ & constant initial time \\
$t_{ex}$ & constant execution horizon \\
$w_{s0}$ and $w_{f0}$ & weight terms on the slack variables \\
%\onslide<4->{
$x_0$ & desired initial state vector \\
$x_{0_{tol}}$ & initial state tolerance vector \\
$x_f$ & desired final state vector \\
$x_{f_{tol}}$ & final state tolerance vector \\
\hline
\end{tabular}
\end{center}
\end{column}
%}}
\end{columns}

\else
\begin{align}
   & \underset{x(t),u(t),{x_0}_s,{x_f}_s\;t_f}{minimize}
   & {\cal{M}}(x(t_0+t_{ex}),t_0+t_{ex},x(t_f),t_f) & \nonumber \\
   && + \int_{t_0+t_{ex}}^{t_f}L(x(t),u(t),t) \, \text{d}t  & \nonumber \\
   && +  w_{s0}{x_0}_s + w_{sf}{x_f}_s \label{eqn:cost} &\\
   & \text{subject to}
   & \frac{\text{d}x}{\text{d}t}(t) - F(x(t),u(t),t) =0 \label{eqn:dynamics} &\\
   && C(x(t),u(t),t) \le 0  \label{eqn:path}& \\
   &&  x_0-x_{0_{tol}} \leq x(t_0+t_{ex}) \leq x_0+x_{0_{tol}}  \label{eqn:x0}& \\
   &&  x_f-x_{f_{tol}} \leq x(t_f) \leq x_f+x_{f_{tol}}  \label{eqn:xf}&\\
   &&  x_{min} \leq  x(t) \leq  x_{max}  \label{eqn:stboundary}&\\
   &&  u_{min} \leq  u(t) \leq  u_{max}  \label{eqn:ctrboundary}&\\
   &&   t_{f_{min}} \leq  t_f \leq  t_{f_{max}}  \label{eqn:time_constraint} &\\
   && x_0 - x(t_0+t_{ex}) \leq {x_0}_s   \label{eqn:sl1} &\\
   && x_0 + x(t_0+t_{ex}) \geq  {x_0}_s  \label{eqn:sl2} &\\
   && x_f - x(t_f) \leq {x_f}_s  \label{eqn:sl3} &\\
   && x_f + x(t_f) \geq  {x_f}_s  \label{eqn:sl4} &
\end{align}
where $t_0$ is the fixed initial time, $t_{ex}$ is the fixed execution horizon that is added to account for the non-negligible solve-times in NMPC applications, $t_f$ is the free final time, $t$ is the time, $x(t) \in \R^{n_{st}}$ is the state, with $n_{st}$ defined as the number of states, and $u(t) \in \R^{n_{ctr}}$ is the control, with $n_{ctr}$ as the number of controls. $x_{s0} \in \R^{n_{st}}$ and $x_{sf} \in \R^{n_{st}}$ are optional slack variables for the initial and terminal states, respectively.
%
%The objective functional includes ${\cal{M}}: \R^{n_{st}} \times \R\times\R^{n_{st}}\times\R\rightarrow\R$, $L:\R^{n_{st}}\times\R^{n_{ctr}}\times\R \rightarrow\R$, and ${\cal{S}}(x_{s0},x_{sf}): \R^{n_{st}} \times \R^{n_{st}} \rightarrow \R$ which are the Mayer, Lagrangian, and slack variable terms, respectively.
%Here ${\cal{S}}(x_{s0},x_{sf})$ is added to the Bolza form to accommodate slack variables on the initial and terminal conditions, this term is described in detail later in this section.
The objective functional includes ${\cal{M}}: \R^{n_{st}} \times \R\times\R^{n_{st}}\times\R\rightarrow\R$ and $L:\R^{n_{st}}\times\R^{n_{ctr}}\times\R \rightarrow\R$, which are the Mayer and Lagrangian terms, respectively.
Here $w_{s0}{x_0}_s + w_{sf}{x_f}_s$ is added to the Bolza form to accommodate slack variables on the initial and terminal conditions; this term is described in detail later in this section.
\Huck{change to weight variables}
${x_0}_s \in \R^{n_{st}}$ and ${x_f}_s \in \R^{n_{st}}$ are vectors of weight terms on the slack variables for the initial and final state constraints.
$F:\R^{n_{st}}\times \R^{n_{ctr}} \times \R \rightarrow \R^{n_{st}}$ and $C:\R^{n_{st}}\times\R^{n_{ctr}} \times \R \rightarrow\R^p$ denote the dynamic constraints and the path constraints, respectively; $p$ is the number of path constraints.
$x_0\in \R^{n_{st}}$ and $x_f\in \R^{n_{st}}$ denote the desired initial and final states, respectively. $x_{0_{tol}} \in \R^{n_{st}}$ and $x_{f_{tol}} \in \R^{n_{st}}$ establish tolerances on the initial and final state, respectively.
Constant upper and lower bounds on the state, control, and final time are included with Eqn. \ref{eqn:stboundary}, Eqn. \ref{eqn:ctrboundary}, and Eqn. \ref{eqn:time_constraint}, respectively.
Finally, \NLOpt adds Eqn. \ref{eqn:sl1} - Eqn. \ref{eqn:sl4} to the Bolza form for optional slack constraints on the initial and terminal states.
\fi

%%%%%%%%%%%
\NLOpt is embedded in the \Julia language and specializes \JuMPnsp's syntax to better suit the domain of optimal control.
\JuMP leverages \Juliansp's syntactic macros \cite{bezanson2012julia} to enable a natural algebraic syntax for modeling optimization problems, without sacrificing performance or restricting problem formulations \cite{LubinDunningIJOC}.
\NLOpt extends \JuMP to include syntax for modeling OCPs in Boltza form in Eqn. \ref{eqn:cost} -- Eqn. \ref{eqn:time_constraint}, with the option of including slack constraints on the initial and terminal states through Eqn. \label{eqn:sl1} -- Eqn. \label{eqn:sl4}.

For a basic example of this syntax, \NLOpt is now used to model the Bryson-Denham problem, which is given in mathematical form as
\begin{equation*}
  \begin{aligned}
     & \underset{a(t)}{\textbf{minimize}}
     & & \frac{1}{2}\int_0^{1}a(t)^2dt\\
     & \textbf{subject to} \\
       \dot{v}(t) =& a(t), &  \dot{x}(t) =& v(t),& x(t) \leq \frac{1}{12} &\\
   &        v(0)=&-v(1)=1, & x(0)       =&x(1)=0 &\label{eq:bd}\\
  \end{aligned}
\end{equation*}

The \lstinline!define()! function is used to create a model object and define Eqn. \ref{eqn:x0} - Eqn. \ref{eqn:ctrboundary} as
\begin{lstlisting}
n = define(numStates = 2, numControls = 1, X0 = [0.,1.], XF = [0.,-1.], XL = [0.,NaN], XU = [1/12,NaN], CL = [NaN,NaN], CU = [NaN,NaN])
\end{lstlisting}
where \lstinline!n! is an object that holds the entire optimal control model, \lstinline!numStates! and \lstinline!numControls! are the number of states and controls, \lstinline!X0! and \lstinline!XF! are arrays of the initial and final state constraint, \lstinline!XL! and \lstinline!XU! are arrays of any lower and upper state bounds, \lstinline!NaN! indicates that a particular constraint is not applied, and \lstinline!CL! and \lstinline!CU! are an arrays of any lower and upper control bounds.

The dynamic constraints in Eqn. \ref{eqn:dynamics} are then added to the model through the \lstinline!dynamics! function as
\begin{lstlisting}
dynamics!(n, [:(x2[j]), :(u1[j])])
\end{lstlisting}
where the ! character indicates that the model object \lstinline!n! is being modified by the function. The elements of the array \lstinline!:(x2[j])! and \lstinline!:(u1[j])! represent $v(t)$ and $a(t)$; by default the state and control variables are \lstinline!x1,x2,..! and \lstinline!u1,u2,..!, but they can be changed. Differential equations must be passed within an array of  \Julia expressions (i.e., \lstinline![:(),:(),...,:()]!), and the index \lstinline![j]! must be appended to the state and control variables. \lstinline!j! is used within \NLOpt to index particular time discretization points $\in[t_0+t_{ex},t_f]$.

The next step is to indicate whether or not the final time $t_f$ is a design variable using the \lstinline!configure! function as
\begin{lstlisting}
configure!(n; (:finalTimeDV => true))
\end{lstlisting}
where \lstinline!(:finalTimeDV=>true)! indicates that the final time is a design variable, which is the case for the Bryson-Denham problem. Additional options can be passed to the \lstinline!configure! function. However, this paper is not a tutorial; for a tutorial see \NLOptnsp's documentation \cite{febbo_2017}.

At this point, any path constraints in  Eqn. \ref{eqn:path} can be added to the model using \JuMPnsp's \lstinline!@NLconstraint! macro. However, these constraints are not needed for this example.

Next, the objective function in Eqn. \ref{eqn:cost} is added to the model.
To accommodate for a Lagrangian term, \NLOpt provides the \lstinline!integrate! function---similar to the \lstinline!dynamics! function, an expression must be passed and the \lstinline![j]! syntax must be appended to all state and control variables.
For the Bryson-Denham, the objective functional is modeled as
\begin{lstlisting}
obj = integrate!(n, :(0.5*u1[j]^2))
\end{lstlisting}
The \JuMP macro \lstinline!@NLobjective! is used to add the objective functional to the model as \lstinline!@NLobjective(n.ocp.mdl,Min,obj)!.
This problem is solved by passing the model \lstinline!n! to the \lstinline!optimize! function as
\begin{lstlisting}
optimize!(n)
\end{lstlisting}

\begin{figure}
\begin{center}
\includegraphics[width=\linewidth]{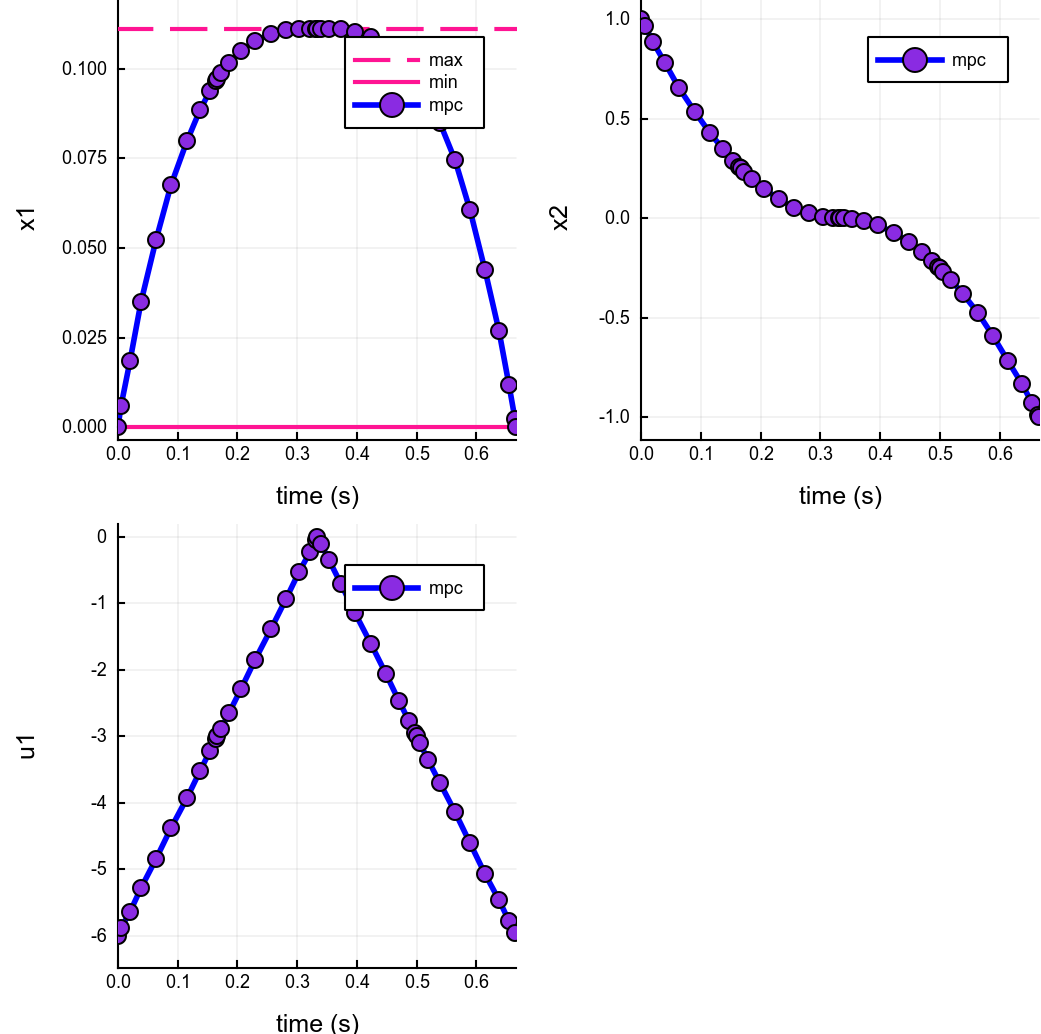}
\caption{Output of \lstinline!allPlots(n)! command after modeling and solving the Bryson-Denham problem using \NLOptnsp. Section~\ref{sec:bryson} in the Appendices provides additional plots of the \NLOptnsp's solution to the Bryson-Denham problem compared to the analytical solution, including the costates.}\label{fig:bdplot}
\end{center}
\end{figure}

\paragraph{Visualization}
\NLOpt allows users quickly plot the solutions to their problems. For plotting ---by default--- \NLOpt leverages \GR \cite{Heien_2018} as a backend, but it can be configured to utilize \matplotlib \cite{Hunter_2018} instead. The command \lstinline!allPlots(n)! plots the solution trajectories for the states, controls, and costates\footnote{if \lstinline!n.s.ocp.evalCostates! is set to \lstinline!true!}. Invoking this command to visualize the solution to the Bryson-Denham problem that is modeled above produces Fig. \ref{fig:bdplot}.

\subsection{Nonlinear model predictive control}
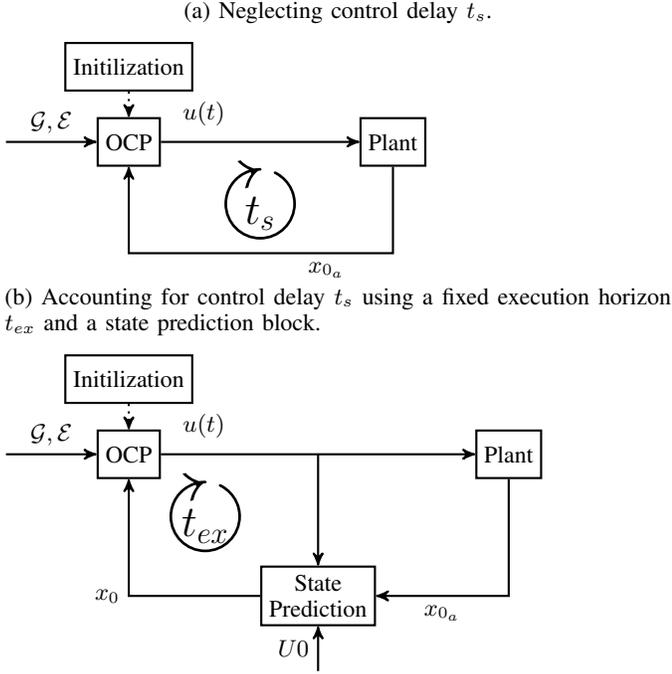
\begin{figure}
\centering
\begin{subfigure}[b]{\linewidth}
\caption{Neglecting control delay $t_s$.}
\begin{tikzpicture}[scale=1, auto, >=stealth']
\small
\lineskip -2pt
 \matrix[ampersand replacement=\&, row sep=1.15cm, column sep=1.32cm]{
 \node [block, name=OCP] {OCP}; \& \node [output] (u1) {}; \& \node [block, name=V] {Plant};  \\
 \node [output] (q1) {}; \& \node [output,name=E] {};  \& \node [output] (q2) {}; \\
                                                                         };
\node [output, left of=OCP] (IN) {};
\node [output, below of=E] (ui) {};
\huge
\node [above of=E, label={[shift={(0,-1.3)}]\CircArrowRight{$t_{s}$}}] {};
\small
\node [right of=OCP, label={[shift={(0,0)}]$u(t)$}] {};
\node [output, right of=E] (RE) {};
\node [output, left of=E] (LE) {};
\node [output, above of=V] (iii) {};

    \node [block, above of=OCP] (I) {Initilization};
    \draw [connector,dotted] (I) -- (OCP);
    \node [output, below of=OCP] (A) {};
    \node [output, below of=V] (B) {};
   %  \draw [connector] ([xshift=0em]ui) -- node[name=uo] {$U0$} ([xshift=0em]E);
    \draw [connector] ([xshift=-2em]IN.west) -- node[name=L] {$\cal{G},\cal{E}$} (OCP);
   % \draw [connector] (iii) -- node {$X0,U0$} (V);
    \draw [connector] (OCP) -- (V);
    \draw [connector] (V) -- (q2) -- node[name=XA] {$x_{0_a}$} (E)-- (q1) -| (OCP);
   % \draw [connector] (E) -- (q1) -| node[name=XB] {} (OCP);
 %   \draw [connector] ([xshift=0cm]u1) -- node[name=UB] {}  ([xshift=0cm]E);
\end{tikzpicture}
   \label{fig:NA}
\end{subfigure}

\begin{subfigure}[b]{\linewidth}
   \caption{Accounting for control delay $t_s$ using a fixed execution horizon $t_{ex}$ and a state prediction block.}
\begin{tikzpicture}[scale=1, auto, >=stealth']
\small
\lineskip -2pt
 \matrix[ampersand replacement=\&, row sep=1.15cm, column sep=1.32cm]{
 \node [block, name=OCP] {OCP}; \& \node [output] (u1) {}; \& \node [block, name=V] {Plant};  \\
 \node [output] (q1) {}; \& \node [block, name=E] { State \\ Prediction };  \& \node [output] (q2) {}; \\
                                                                         };
\node [output, left of=OCP] (IN) {};
\node [output, below of=E] (ui) {};
\huge
\node [above of=E, label={[shift={(-1.5,-0.9)}]\CircArrowRight{$t_{ex}$}}] {};
\small
\node [right of=OCP, label={[shift={(0,0)}]$u(t)$}] {};
\node [output, right of=E] (RE) {};
\node [output, left of=E] (LE) {};
\node [output, above of=V] (iii) {};

    \node [block, above of=OCP] (I) {Initilization};
    \draw [connector,dotted] (I) -- (OCP);
    \node [output, below of=OCP] (A) {};
    \node [output, below of=V] (B) {};
     \draw [connector] ([xshift=0em]ui) -- node[name=uo] {$U0$} ([xshift=0em]E);
    \draw [connector] ([xshift=-2em]IN.west) -- node[name=L] {$\cal{G},\cal{E}$} (OCP);
   % \draw [connector] (iii) -- node {$X0,U0$} (V);
    \draw [connector] (OCP) -- (V);
    \draw [connector] (V) -- (q2) -- node[name=XA] {$x_{0_a}$} (E);
    \draw [connector] (E) -- (q1) -| node[name=XB] {$x_0$} (OCP);
    \draw [connector] ([xshift=0cm]u1) -- node[name=UB] {}  ([xshift=0cm]E);
\end{tikzpicture}

   \label{fig:NB}
\end{subfigure}
\caption{Nonlinear model predictive control framework available in \NLOptnsp.}\label{fig:mpc}
\end{figure}

%%%%%%%%%%%%%%
Fig. \ref{fig:mpc} depicts two ways that \NLOpt can be used to solve OCPs for NMPC applications; Fig. \ref{fig:NA} neglects control delays and Fig. \ref{fig:NB} accounts for them.
This section describes these figures and discusses the design features that help \NLOpt users tackle NMPC problems.
%
%These features permit the user to
%\begin{itemize}
%\item modify the constraint tolerances on the initial and terminal states,
%\item include slack constraints,
%  fix .. design framework?
%\item and initialize the OCP
%\end{itemize}

Fig. \ref{fig:NA} has three main components: the OCP, the plant, and the initialization block.
% TODO consider putting xf as an input..in addition to G or constraints on xf or something.
Three inputs $\cal{G}$, $\cal{E}$, and $x_0$ are provided to the OCP to produce $u(t)$; $u(t)$ can be either a reference trajectory or control signals for the plant.
In the case that $u(t)$ is a reference trajectory, then low-level controllers are added to the plant to allow it to track the trajectory.
%In the example provided in this work, since there are no low level-controllers in this work, u(t) is the control signal.
%

\begin{descrip}{Goal information $\cal{G}$}
includes the final desired state of the plant, which may not be equal to $x_f$. For instance, in an automated vehicle trajectory planning system, the goal range may be outside of the sensing range. In this case, the final desired state $x_f$ may be near the boundary of the sensing range.
\end{descrip}
\begin{descrip}{Environment information $\cal{E}$}
includes any transient data. For example, this data may include the obstacle data that helps establish the constraints on obstacle avoidance for automated vehicle navigation problems.
\end{descrip}

%%%%%%%%%%
The plant can be either physical or virtual, but in either case is provided by the user.
Because time can typically be allocated to initialize NMPC problems, the initialization block permits users to warm start their optimization problems so that the initial on-line solve time is much smaller.
After initialization, at $t_0$, the first control signal $u(t)$ is sent to the plant and the first on-line OCP is solved.
Each time an OCP-solve starts, $t_0$ is reset to the current time.
An issue with this scheme is that it does not take into account the solve time (i.e., control delay $t_s$).
That is, the initial state of the OCP is constrained to be the current state of the plant $x_0$ at the initial time $t_0$, so by the time the OCP has been solved $t_s$ has elapsed, and the plant will have evolved to a new state.
If this control delay is small relative to the time scale of the dynamics, then neglecting it will not compromise the robustness.
However, if the control delay is relatively large, then it cannot be neglected.
%

%%%%%%%%%%%%%%% TODO talk about U0
Fig. \ref{fig:NB} illustrates an approach that accounts for these control delays.
This approach adds a block that predicts the plant state at the current time plus a fixed execution horizon $t_0+t_{ex}$.
The execution horizon $t_{ex}$ can be chosen based on the a heuristic upper limit on the solve times; often solve times do not change drastically when solved in a receding-horizon with varying parameters for the initial conditions and path constraints.
This approach avoids having to predict individual solve times $t_s$.

%%%%%%%%%%%%%
%\NLOpt provides (1) functionality to help avoid infeasibilities when closing the loop and (2) A design framework to help the user formulate their OCP for use in NMPC applications.
%\NLOpt provides functionality to help avoid infeasibilities when closing the loop in NMPC applications.
\NLOpt provides various functionality tailored for solving NMPC problems.
% TODO make consistent with list above..OR do not repeat!
%
The remainder of this section simultaneously describes these features and provides an example that uses \NLOpt to formulate an OCP and solve it in a receding horizon.
To this end, consider the moon lander OCP \cite{meditch1964problem}, which is given in without slack constraints in Eqn. \ref{eq:ml} as
\begin{equation}
  \begin{aligned}
     & \underset{a(t),\;t_f}{\text{minimize}}
     & \int_0^{t_f} a(t) \text{dt} \\
     & \text{subject to}
     &     \dot{x}(t) = v(t),     &&\dot{v}(t) = a(t) -g\\
     & &    x(t_0)     =10,        &&   x(t_f)=0\\
     & &    v(t_0)     =-2,         &&  v(t_f)=0\\
     & &   0     \leq x(t) \leq 20, && -20 \leq v(t) \leq 20  \\
     & &   0       \leq a(t) \leq 3, && 0.001 \leq t_f \leq 400 \label{eq:ml} \\
  \end{aligned}
\end{equation}
where the $x(t)$ is the altitude, $v(t)$ is the speed, $a(t)$ is the thrust, $g=1.5$ is the local gravitational acceleration, $t_f$ is the final time. The objective is to minimize the thrust of the spaceship given the dynamic constraints, event constraints, control constraints, and final time constraints.

Listing. \ref{lst:moonMPC} shows the code needed to solve Eqn. \ref{eq:ml} as an MPC problem.
Line $1$ creates a model \lstinline!n! of the OCP with the initial and terminal state constraints, and the constant upper and lower bounds on the state and control variables.
As is, the model \lstinline!n! has low-tolerance hard constraints on the initial and terminal state conditions.
However, these low-tolerance hard constraints can lead to infeasible problems and longer solve-times, especially when the loop is closed.
That is, when the control drives the plant into an infeasible state space, an infeasible problem is engendered \cite{scokaert1999feasibility}; and typically, the solve-times increase as the problems become less-feasible.
Therefore, an ability to easily adjust these low-tolerance constraints to high-tolerance hard constraints is desirable.
%As a first step to avoiding these
As seen in Line $2$, \NLOpt enables this feature through the \lstinline!defineTolerances! function.
\lstinline!X0_tol! and \lstinline!XF_tol! are arrays that set the tolerances on the initial $x_{0_{tol}}$ and final states $x_{f_{tol}}$, respectively.
%
%Line $6$ adds the dynamic constraints.
%
\if\slides1
\begin{minipage}{\linewidth}
\begin{lstlisting}[caption={\NLOpt code needed to formulate and solve the moon lander problem.},label=lst:nl_moon,frame=lines]
n = define(numStates = 2, numControls = 1, X0 = [10, -2], XF = [0., 0.], XL = [0, -20], XU = [20, 20], CL = [0.], CU = [3.]);
dynamics!(n, [:(x2[j]), :(u1[j] - 1.5)]);
configure!(n; (:finalTimeDV => true));
obj = integrate!(n, :(u1[j]));
@NLobjective(n.ocp.mdl, Min, obj);
optimize!(n);
\end{lstlisting}
\end{minipage}
\else
\begin{figure*}
\begin{lstlisting}[caption={\NLOpt code needed to formulate and solve the moon lander as an MPC problem.},label=lst:moonMPC,frame=lines,    basicstyle = \small\ttfamily]
n = define(numStates = 2, numControls = 1, X0 = [10., -2], XF = [0., 0.], CL = [0.], CU = [3.])
defineTolerances!(n; X0_tol = [0.01, 0.005], XF_tol = [0.01, 0.005])
dynamics!(n,[:(x2[j]),:(u1[j]-1.5)])
configure!(n; (:finalTimeDV => true), (:xFslackVariables => true), (:x0slackVariables => true))
obj = integrate!(n,:(u1[j]))
@NLobjective(n.ocp.mdl, Min, obj + 100*(n.ocp.x0s[1] + n.ocp.x0s[2] + n.ocp.xFs[1] + n.ocp.xFs[2]))
initOpt!(n)
defineMPC!(n; tex = 0.2, predictX0 = true)
function IPplant(n, x0, t, U, t0, tf)
  spU = linearSpline(t, U[:,1])
  f = (dx, x, p, t) -> begin
    dx[1] = x[2]
    dx[2] = spU[t] - 1.5
  end
  return DiffEqBase.solve(ODEProblem(f, x0, (t0, tf)), Tsit5()), [spU]
end
defineIP!(n, IPplant)
simMPC!(n)
\end{lstlisting}
%\caption{The ! character on the functions means that the model is being modified.}
\end{figure*}
\fi

%dx = [:(x2[j]),:(u1[j]-1.5)]
% where, by default, when the final time is a design variable it is constrained to be between $0.001$ and $1000$

When going from low- to high-tolerance hard constraints on the initial and terminal states, slack constraints should also be added.
Because, when using these high-tolerance constraints without slack constraints, there is nothing pushing the initial and terminal states away from the edge of the infeasible region.
Thus, infeasible problems are just as likely to occur.
Adding slack constraints on the initial and terminal state constraints helps to mitigate these infeasible problems.
Before slack constraints are added to the model, slack variables must be added.
The size of a slack variable corresponds to the size of the respective constraint violation \cite{kerrigan2000soft}.
As seen in Line $4$, \NLOpt allows such slack variable to be added using the \lstinline!configure! function.
\lstinline!(:xFslackVariables=>true)! and \lstinline!(:x0slackVariables=>true)! adds slack variables on the initial and final state constraint, respectively.
Both the objective of the moon lander problem and the slack constraints are added to model as on Line $6$.
\lstinline!n.ocp.x0s! and \lstinline!n.ocp.xFs! are arrays holding the slack variables on the initial and terminal states, respectively, and all of the terms in $w_{s0}$ and $w_{sf}$ (in Eqn. \ref{eqn:cost}) are set to $100$---these weights are set large enough such that the respective constraint violations are nearly zero.
On Line $7$, \NLOpt warm starts the optimization using the  \lstinline!initOpt! function; the initialization block in Fig. \ref{fig:mpc} captures this step.

The \lstinline!defineMPC! function adds several basic settings to the model \lstinline!n!.
\lstinline!tex! is the value of the fixed execution horizon and \lstinline!predictX0! is a bool, which, when set to \lstinline!true!, indicates that the the framework in Fig. \ref{fig:NB} is used. Thus, a prediction of the initial state needs to be made either by the user or using an internal model of the plant, which is added to \lstinline!n!.
In this simple example, the differential equations in Eqn. \ref{eq:ml} govern the OCP, the plant, and the state prediction function.
The plant and prediction model are defined by the \lstinline!IPplant! function from Line $9$ to Line $16$ and passed to the model \lstinline!n! using the \lstinline!defineIP! function on Line $17$.
Here the \lstinline!IPplant! function is showed for completeness, but its is not described in detail since it uses the well-documented \DifferentialEquations package in \Julia \cite{rackauckas2017differentialequations}.
For safety and reduced time in experimental development, this initial step, i.e., making all of the models the same and running a simulation-based experiment, should be taken; especially when formulating more complex OCPs for practical NMPC applications.

\begin{figure}
\centering
\begin{subfigure}[b]{\linewidth}
\caption{Output of \lstinline!mpcPlots(n,4)! command. }\label{fig:MLA}
\includegraphics[width=\linewidth]{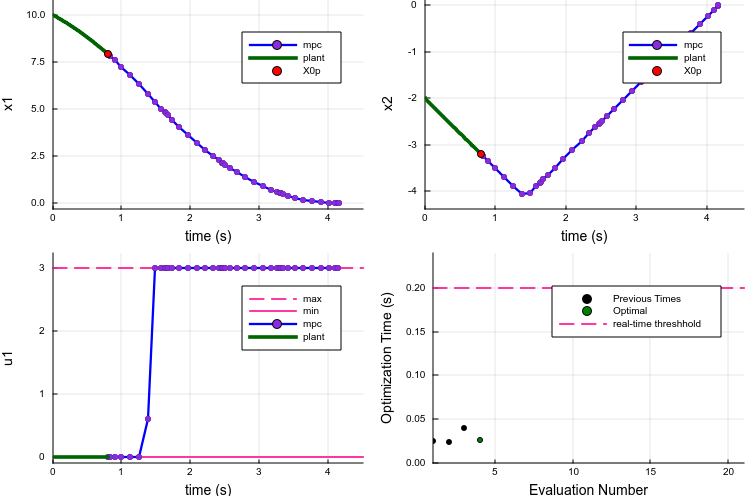}
\end{subfigure}

\begin{subfigure}[b]{\linewidth}
\caption{Output of \lstinline!mpcPlots(n,15)! command.}\label{fig:MLB}
\includegraphics[width=\linewidth]{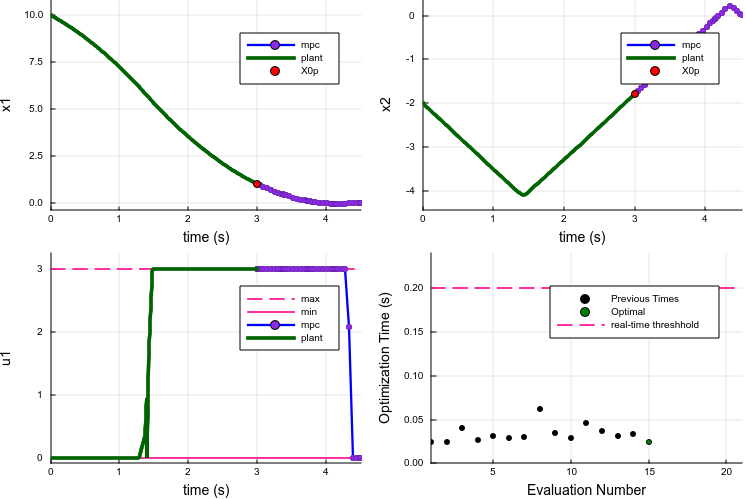}
\end{subfigure}

\caption{Closed-loop visualization of moon lander problem using \NLOptnsp.}
\end{figure}

\paragraph{Visualization}
The command \lstinline!mpcPlots(n,idx)! plots the data for both plant and solution trajectories for the states and controls, the predicted initial state, and the optimization times, where \lstinline!idx! is an integer representing the iteration number. For instance, invoking this command to visualize data at the $4$th and $15$th iterations of the moon lander problem produces Fig. \ref{fig:MLA} and Fig. \ref{fig:MLB}, respectively; \NLOpt provides visualization functionality to combine the frames from all iterations into a single animation.

Section~\ref{sec:moonCL} in the Appendices provides a  plot \NLOptnsp's closed-loop solution to the moon lander problem compared to the analytical solution.

%
%The closed-loop results of the code in Listing. 1 are shown in Fig. 2.
\section{Numerical optimal control}\label{sec:direct}
This section provides an overview of numerical optimal control methods (i.e., transcription methods).
%
%Then, then various direct-collocation methods is presented.
%
The goal of this section is to motivate the choice of direct-collocation methods in \NLOptnsp, not to provide the reader with a complete description of numerical optimal control methods.
Readers are referred to \cite{betts1998survey,biegler2007overview,Betts2010}, for more comprehensive reviews on this subject.
After these methods are discussed, the mathematics of the various direct-collocation methods as they are implemented in \NLOpt are provided.
%, where all of the these methods are established by previous work [cite].
%
%Thus, this work does not provide a mathematical descriptions of the direct-collocation methods implemented within \NLOpt.
%
%Instead the reader is referred to the seminal work [cite] and this paper focuses on its novel contributions.
% consider introducing some of the math...

\subsection{Numerical optimal control overview}
Tractable exact algorithms for solving OCPs suitable for practical applications do not exist; thus, numerical methods are used \cite{paden2016survey}.
Numerical methods for solving OCPs (i.e., trajectory optimization problems) are generally broken into two categories: indirect and direct methods.
Indirect methods seek the root of the necessary conditions for optimality \cite{pontryagin1987mathematical} while direct methods seek the extrema of the cost functional \cite{betts1998survey}.
Compared to direct methods, indirect methods produce better error estimates \cite{kelly2017introduction} and require less preliminary work to determine optimality \cite{garg2011advances}.
However, indirect methods have several disadvantages: the necessary conditions must be derived \cite{hull1997conversion}; the incorporation of path constraints requires an a priori estimation of the sequence of constrained/unconstrained singular arcs; and a guess needs to be made for the adjoint variables \cite{Betts2010}.
Due to these disadvantages, \NLOpt solves OCPs using direct methods.

Direct methods are broken into shooting methods \cite{osborne1969shooting}, multiple shooting methods \cite{bock1984multiple,diehl2006fast}, and direct-collocation methods.
Shooting methods are not suitable for most practical applications because they do not work well when the number of variables is large \cite{sargent2000optimal}.
The multiple shooting method is well suited to exploit parallel processing due to the structure of its formulation \cite{Betts2010,diehl2002real}.
However, there are several disadvantages to the multiple shooting method: an expensive numerical integration needs to be performed during each iteration of the NLP solve, it can be difficult to incorporate state inequalities \cite{sargent2000optimal}, and using multiple integration steps reduces the sparsity of the Hessian and Jacobian matrices \cite{betts1999exploiting}.
Direct-collocation methods overcome these issues by enforcing the dynamic state constraints within the NLP.
While this results in a larger problem, there is no need to perform expensive integrations of the state dynamics between iterations, because constraints in the NLP enforce the state dynamics at the collocation points, the path constraints can be easily incorporated, and the sparsity in the derivative matrices is preserved.

\subsection{Direct-collocation method overview}
Direct-collocation methods are divided into three categories of polynomial approximation types: h-methods (or local methods) \cite{hargraves1987direct, hager2000runge, betts1998survey}, p-methods (or global methods) \cite{garg2010unified, fahroo2008advances, huntington2007comparison}, and hp-methods (a hybrid of the h- and p-methods) \cite{patterson2015ph, patterson2014gpops, Darby2011}.
In an h-method, the dynamic state constraints are satisfied using local approximations; e.g., Euler's method or the trapezoidal method \cite{hargraves1987direct}.
For h-methods, increasing the number and location of the collocation points \cite{Betts2010, patterson2014gpops, jain2008trajectory} leads to convergence.
However, a large number of points may be required for convergence, which can result in large solve-times \cite{darby2009mesh}.
p-methods can reduce the number of points needed for convergence, because they are more accurate than h-methods \cite{elnagar1995pseudospectral}.
p-methods approximate OCPs using global polynomials constructed by collocating the dynamics at Gaussian quadrature points \cite{elnagar1995pseudospectral}.
p-methods were originally developed to solve problems in computational fluid dynamics \cite{hussaini1987spectral} and since have been used in practice in optimal control.
% todo: this is the textbook take-home, what is the main message.. traditional topic sentence
%Since then, p-method's have been used in practice in optimal control.
%
For instance, p-methods were used to rotate the International Space Station $180$ degrees without using any propellant\footnote{The cost of the fuel saved was estimated at one million dollars and control of the space station orientation was accomplished using gyroscopes \cite{kang2007pseudospectral}. } \cite{fahroo2008advances}.
%
%When compared to h-methods, while p-methods can reduce the number of collocation points needed for convergence \cite{darby2009mesh}, the Jacobian and Hessian matrices are much denser, which results in a larger NLP \cite{darby2011hp}.
A drawback with p-methods is that the Jacobian and Hessian matrices are much denser than with h-methods, which results in a larger NLP \cite{darby2011hp}.

By construction, hp-methods help to mitigate the accuracy issues with h-methods and the NLP problem size with p-methods.
Instead of using a single polynomial as with p-methods, hp-methods use multiple polynomials constrained to be connected to one another at the endpoints. This construction reduces the size of the NLP while maintaining accurate approximations \cite{Darby2011}.
\subsection{Direct-collocation methods in \NLOpt}
At the time of this writing, three direct-collocation methods are implemented in \NLOptnsp: two h-methods and one p/hp-method.
The remainder of this section illustrates how these methods are implemented in \NLOptnsp.

\subsubsection{h-Methods}
%h-methods (also known as time marching methods) make local approximations of functions, which results in very sparse Jacobian and Hessian matrices within the NLP \cite{Darby2011}.
Euler's backward method and the trapezoidal method are embedded in \NLOptnsp.
However, before these h-methods are given, the h-discretization matrices used to approximate the continuous-time OCP are provided.

\paragraph{h-Discretization Matrices}
Consider that $t$ is sampled at $N$ evenly spaced discritization points $\in [t_0+t_{ex},t_f]$ and denote the result as the vector $\mathbf{T}=[T_1,\dots,T_N]$. Then, for instance the $t_0+t_{ex}$ and $t_f$ are defined as $T_1$ and $T_N$, respectively.
Denote the state and control discretization matrices as
$$x(t) \Big|_{t=\mathbf{T}} = \mathbf{X}$$
and
$$u(t) \Big|_{t=\mathbf{T}} = \mathbf{U}\text{,}$$
respectively. $\mathbf{X}[i]$ is the state at the $i^{th}$ collocation point; thus, $\mathbf{X}[1]$ and $\mathbf{X}[N]$ index the values of the initial and final states, respectively.
The control matrix is similarly defined; $\mathbf{U}[i]$ is the control at the $i^{th}$ collocation point.
Denote the minimum and maximum discretized state limit matrices as
$$x_{min}(t)\Big|_{t=\mathbf{T}} = \mathbf{X}_{min}$$
and
$$x_{max}(t)\Big|_{t=\mathbf{T}} = \mathbf{X}_{max}\text{,}$$ respectively. Similarly, the minimum and maximum control limit matrices are denoted as
$$u_{min}(t)\Big|_{t=\mathbf{T}} = \mathbf{U}_{min}$$
and
$$u_{max}({t})\Big|_{t=\mathbf{T}} = \mathbf{U}_{max}\text{,}$$ respectively.

\paragraph{Euler's Backward Method}
The dynamic constraints in Eqn. \ref{eqn:dynamics} are locally approximated at $(N-1)$ points defined by $\mathbf{T}[2:N]$. To accomplish this, $(N-1)\times n_{st}$ implicit constraints are added as shown in Eqn. \ref{eqn:euler_dyn}
\begin{align}
\mathbf{0} &= \mathbf{X}[i+1] - \mathbf{X}[i] - hF(\mathbf{X}[i+1],\mathbf{U}[i+1],\mathbf{T}[i+1]) \label{eqn:euler_dyn} \\
&=\boldsymbol{\eta}_i,\;\mathrm{for}\;i \in (1:N-1) \nonumber
\end{align}
\benoit{Is this defined ? (1:N-1)}
where $h$ is the time-step size, which is determined by dividing the time span ($t_f-t_0-t_{ex}$) by $N$.

The integral term in the cost functional in Eqn. \ref{eqn:cost} is approximated in Eqn. \ref{eqn:euler_cost} as
\begin{align}
  I=h\sum_{i=1}^{N} L(\mathbf{X}[i],\mathbf{U}[i],T_i) \label{eqn:euler_cost}
\end{align}

\paragraph{Trapezoidal Method}
Similar to Euler's backward method, the dynamic constraints in Eqn. \ref{eqn:dynamics} are locally approximated at $(N-1)$ points defined by $\mathbf{T}[2:N]$. To accomplish this, the $(N-1)\times n_{st}$ implicit constraints in Eqn. \ref{eqn:trapezoidal_dyn} are enforced with
\begin{align}
\mathbf{0} &= \mathbf{X}[i+1] - \mathbf{X}[i]- \frac{h}{2}( F(\mathbf{X}[i],\mathbf{U}[i],\mathbf{T}[i]) + \nonumber \\
           & F(\mathbf{X}[i+1],\mathbf{U}[i+1],\mathbf{T}[i+1])) \label{eqn:trapezoidal_dyn}
\\&=\boldsymbol{\eta}_i,\;\mathrm{for}\; i \in (1:N-1)  \nonumber
\end{align}
Next, the integral term in the cost functional in Eqn. \ref{eqn:cost} is approximated in Eqn. \ref{eqn:trapezoidal_cost} as
\begin{align}
I=\frac{h}{2}\sum_{i=1}^{N} (L(\mathbf{X}[i],\mathbf{U}[i],T_i)+L(\mathbf{X}[i+1],\mathbf{U}[i+1],T_{i+1}))\label{eqn:trapezoidal_cost}
\end{align}

\paragraph{Discrete OCP}
The h-method-based discrete OCP is given as
\begin{align}
     & \underset{\mathbf{X},\;{\mathbf{U}},\;T_N}{\textbf{minimize}}
     && {\cal{M}}(\mathbf{X}[1],T_1,\mathbf{X}[N],T_N) + I \label{eq:costd_tm}\\
     & \textbf{subject to}
     && \boldsymbol{\eta}=\mathbf{0}
   \label{eq:dynamicsd_tm}\\
     &&& \mathbf{C}(\mathbf{X},\mathbf{U},\mathbf{T}) \le \mathbf{0} \label{eq:pathd_tm}\\
     &&& {\boldsymbol{\phi}}(\mathbf{X}[1],T_1,\mathbf{X}[N],T_N)=\mathbf{0} \label{eq:eventd_tm}\\
     &&&  \mathbf{X}_{min} \leq \mathbf{X} \leq \mathbf{X}_{max} \\
     &&&  \mathbf{U}_{min} \leq  \mathbf{U} \leq \mathbf{U}_{max} \\
     &&&  t_{f_{min}} \leq  T_N \leq  t_{f_{max}} \label{eq:time_constraintd_tm}
\end{align}
where slack constraints can be included with the Mayer term in Eqn. \ref{eq:costd_tm} and Eqn. \ref{eq:pathd_tm}.
\subsubsection{{p-Methods} \label{sec:Ps}}
For generality, this paper only describes hp-methods, since the single interval method (i.e., p-method) is merely the case where the number of intervals is equal to one.

\subsubsection{{hp-Methods} \label{sec:miPs}}
The form of Eqn. \ref{eqn:cost} -- Eqn. \ref{eqn:time_constraint} must be modified to directly transcribe the OCP into an NLP using hp-methods.
To apply Gaussian quadrature the interval of integration must be transformed from $[t_0+t_{ex},t_f]$ to $[-1,+1]$.
To accomplish this, $\tau$ $\in$ $[-1,+1]$ is introduced as a new independent variable and a change of variable, for $t$ in terms of $\tau$ using the affine transformation, $ t = \frac{t_f - t_0-t_{ex}}{2}\tau + \frac{t_f + t_0+t_{ex}}{2}$.
Then, the interval $\tau$ $\in$ $[-1,+1]$ is divided into a mesh of $K$ intervals to accommodate for multiple intervals.
With this, as in \cite{darby2009mesh}, an array of mesh points $(M_0,\dots,M_K)$ for the boundaries of these intervals is defined, which satisfy
$$-1 = M_0 < M_1 < M_2  < \dots < M_{K-1} < M_K = 1$$
Denote the continuous-time variables for the state and control are on each mesh interval, $k\in(1,\dots,K)$, by the arrays $x^{(k)}(\tau)$ and $u^{(k)}(\tau)$, respectively.
Next, denote arrays of continuous-time variables for both the minimum and maximum state and control limits on each mesh interval, $k\in(1,\dots,K)$, as $x^{{(k)}}_{min}$, $x^{{(k)}}_{max}$, $u^{{(k)}}_{min}$, and $u^{{(k)}}_{max}$, respectively.
The state continuity between the mesh intervals is ensured with the constraint $x^{(k)}(M_k)=x^{(k+1)}(M_k)$ for $k = (1,\dots,K-1)$ \cite{Darby2011}.
Similar to \cite{patterson2014gpops}, this constraint is enforced programatically by making $x^{(k)}(M_k)$ be the same variable as $x^{(k+1)}(M_k)$.
To continue to describe the hp-method implemented in \NLOpt, the hp-discretization matrices are defined, which hold the discrete-time values of the approximation to continuous-time problem.

\paragraph{hp-Discretization Matrices}
First an array of time discretization vectors, $\boldsymbol{\tau}^{(k)}=[\tau_1^k,\dots,\tau_{N^k}^{k}]$, is defined by evaluating the continuous functions at $N^k$ specified $\tau$'s $\in[M_{k-1},M_{k})$ for $k\in[1,\dots,K]$, where $N^k$ notates the number of collocation points in mesh interval $k$; for instance, $\tau_1^1=-1$.
Let $$\mathbf{N}=[N_1,N_2,\dots,N_k,\dots,N_{K-1},N_K]$$ denote an array that holds the number of collocation points within each mesh interval, where $N^k$ can be adjusted according to the desired level of fidelity for the $k^{th}$ mesh interval.
For $k\in[1,\dots,K]$, denote the state and control discretization matrix arrays as
$$x^{(k)}(\tau) \Big|_{\tau =\boldsymbol{\tau}^{(k)}} = \mathbf{X}^{(k)}$$ and
$$u^{(k)}(\tau) \Big|_{\tau=\boldsymbol{\tau}^{(k)}} = \mathbf{U}^{(k)}\text{,}$$
respectively. Next, denote the minimum and maximum discretized state limit matrix arrays as
$$x^{{(k)}}_{min}(\tau)\Big|_{\tau=\boldsymbol{\tau}^{(k)}} = \mathbf{X}^{{(k)}}_{min}$$
and
$$x^{{(k)}}_{max}(\tau)\Big|_{\tau=\boldsymbol{\tau}^{(k)}} = \mathbf{X}^{(k)}_{max}\text{,}$$
respectively.
Similarly, the minimum and maximum control limit matrices are defined as
$$u^{{(k)}}_{min}(\tau)\Big|_{\tau=\boldsymbol{\tau}^{(k)}} = \mathbf{U}^{{(k)}}_{min}$$
and
$$u^{{(k)}}_{max}(\tau)\Big|_{\tau=\boldsymbol{\tau}^{(k)}} = \mathbf{U}^{{(k)}}_{max}\text{,}$$
respectively. \\

To approximate the modified OCP that is modified for hp-methods, \NLOpt builds on the work done in \cite{garg2010unified,garg2011pseudospectral,patterson2012exploiting}, which was implemented in \GPOPSii \cite{patterson2014gpops}. Specifically, \NLOpt implements the Legendre-Gauss-Radau quadrature collocation method (Radau collocation method). For completeness, this section will briefly describe this method, but for a more thorough explanation, the reader is referred to the seminal work done in  \cite{patterson2014gpops,garg2010unified,garg2011pseudospectral,patterson2012exploiting}.

\paragraph{Radau Collocation Method}
In hp-methods, the states are approximated within each mesh interval with a Lagrange polynomial as

\begin{equation}
x^{(k)}(\tau) \approx \displaystyle\sum_{j=1}^{N^k+1}\mathbf{X}[j]^{(k)}\mathcal{L}^{(k)}_j(\tau),\;k \in [1,..,K]
\end{equation}
with
\begin{equation}
\mathcal{L}_j^k(\tau)=\prod_{\substack{l=1 \\ l\neq j}}^{N^k+1}\frac{\tau-\tau_l^k}{\tau_j^k-\tau_l^k},\;k \in [1,..,K]
\end{equation}
$\mathcal{L}_j^k(\tau)$ is the $(k^{th}, j^{th})$ Lagrange polynomial within a basis of Lagrange polynomials defined by $j=(1,\dots,N^k+1)$ and $k = (1,\dots,K)$, $\boldsymbol{\tau}^{(k)}=[\tau_1^k,\dots,\tau_{N^k}^{k}]$ and is the $k^{th}$ set of the LGR collocation points (also, called LGR nodes \cite{abramowitz1965stegun}), which are defined on the $k^{th}$ mesh interval ($\tau\in[M_{k-1},M_k)$). Then to approximate the entire state, $M_k$ is added as a noncollocated point \cite{garg2010unified} for $k\in(1,\dots,K)$.

%function scale_tau(tau,ta,tb)
 % (tb - ta)/2*tau + (ta + tb)/2;
%end

%function scale_w(w,ta,tb)
 % (tb - ta)/2*w
%end

%function createIntervals!(n::NLOpt)
 %   tm = Ranges.linspace(-1,1,n.Ni+1)     # create mesh points
  %  di = 2/n.Ni                           # interval size
   % # go through each mesh interval creating time intervals; map [tm[i-1],tm[i]] --> [-1,1]
   % n.ts = [[scale_tau(n.tau[int],tm[int],tm[int+1]);di*int-1] for int in 1:n.Ni]
   % n.ws = [scale_w(n.w[int],tm[int],tm[int+1]) for int in 1:n.Ni]
   % nothing
%end

% ts is an array of the scaled times to [-1,1]
% ws is an array of the scaled weights

% Integrate
%    @NLexpression(n.mdl, temp[int=1:n.Ni], (n.tf-n.t0)/2*sum(n.ws[int][j]*integral_expr[int][j] for j = 1:n.Nck[int]) );

% Dynamics
%    dynamics_expr[int][:,st]=@NLexpression(n.mdl, [j in 1:n.Nck[int]], sum(n.DMatrix[int][j,i]*x_int[i,st] for i in 1:n.Nck[int]+1) - ((n.tf)/2)*dx[j,st]  )

The derivative of the state can then be approximated for each mesh interval as
\begin{equation}
\frac{\mathrm{d}x^{(k)}(\tau)}{\mathrm{d}\tau} \approx \displaystyle\sum_{j=1}^{N^k+1}\mathbf{X}[j]^{(k)}\frac{\mathrm{d}\mathcal{L}_j^{(k)}(\tau)}{\mathrm{d}\tau},\;k \in [1,..,K]
\end{equation}
with
\begin{equation}
   \frac{\mathrm{d}\mathcal{L}_j^{(k)}(\tau)}{\mathrm{d}\tau}\Big|_{\tau=\tau_j^k} = D_{ij}^k
\end{equation}
where $D_{ij}^k$ is an element of the $N^k \times N^{k + 1}$ Legendre-Gauss-Radau differentiation matrix in the $k^{th}$ mesh interval, as defined in \cite{garg2010unified}.
%For simplicity, the subscripts $ij$ will now be omitted and this entire matrix will be referred to by mesh interval as $\mathbf{D}^{(k)}$.
%

Next, in order to approximate the integral of the Lagrange term in Eqn. \ref{eqn:cost}, Gaussian-Legendre quadrature \cite{gauss1815methodus} is used as
\if\doublecol1
\begin{align}
& \int_{t_0+t_{ex}}^{t_f}L(x(t),u(t),t) \, dt \approx & \nonumber \\
& \frac{t_f-t_0-t_{ex}}{2}\sum_{k=1}^{K} \sum_{j=1}^{N^k} \frac{M_{k}-M_{k-1}}{2} w_j^k L(\mathbf{X}[j]^{(k)},\mathbf{U}[j]^{(k)}, \dots & \nonumber \\
& \tau_j^k;t_0+t_{ex},t_f) & \label{eqn:cost_approx}
\end{align}
\else
\begin{equation}
\int_{t_0+t_{ex}}^{t_f}L(x(t),u(t),t) \, dt \approx \frac{t_f-t_0-t_{ex}}{2}\sum_{k=1}^{K} \sum_{j=1}^{N^k} \frac{M_{k}-M_{k-1}}{2} w_j^k L(\mathbf{X}[j]^{(k)},\mathbf{U}[j]^{(k)}, \tau_j^k;t_0+t_{ex},t_f) \label{eqn:cost_approx}
\end{equation}
\fi
where $\boldsymbol{w}^{(k)} = [w_1^k,\dots,w_{N_k}^{k}]$ is the $k^{th}$ array of LGR weights\footnote{To calculate both the LGR nodes and weights, \NLOpt leverages FastGaussQuadrature \cite{sheehan_2017,abramowitz1965stegun}, which uses methods developed in \cite{hale2013fast}.}.

Eqn. \ref{eqn:cost_approx} is mathematically equivalent to the approximations made for the integral term in the cost functional in \cite{Darby2011}, but it is written in a slightly different form to reduce the computations needed within the NLP. Specifically, the $\frac{M_{k}-M_{k-1}}{2}w_j^k$ term is calculated outside of the NLP, for $j\in(1,\dots,N^k)$ and $k\in(1,\dots,K)$.
The result is stored in an array of vectors. Thus, the design variable $t_f$ is removed from the summations in \NLOptnsp.

\paragraph{Discrete OCP}
%Now that approximations of the modified continuous-time OCP have been made for the dynamics, and the cost functional and the hp-discretized matrices have been defined,
The p-method-based discrete OCP is shown in Eqn. \ref{eq:cost_ps} - Eqn. \ref{eq:time_constraint_ps} as
\if\doublecol1
\begin{align}
   & \underset{\mathbf{X}^{(k)},\;{\boldsymbol{U}^{(k)}},\;t_f}{\textbf{minimize}}
   \; {\cal{M}}(\mathbf{X}[1]^{(1)},t_0+t_{ex},\mathbf{X}[N_{K+1}]^{(K)},t_K)+ I \label{eq:cost_ps}\\
   & \textbf{subject to} & \nonumber \\
  & \sum_{j=1}^{N_k+1}\mathbf{X}_{j}^{(k)}D_{ij}^{(k)}-\frac{t_f-t_0-t_{ex}}{2}\mathbf{f}(\mathbf{X}_{i}^{(k)},\mathbf{U}_{i}^{(k)},\tau_i^k;t_0+t_{ex},t_f)=\mathbf{0} & \\
   & \mathbf{C}^{(k)}(\mathbf{X}[i]^{(k)},\mathbf{U}[i]^{(k)},\tau_{i}^k;t_0+t_{ex},t_f) \le \mathbf{0} &\\
   & \boldsymbol{\phi}(\mathbf{X}[1]^{(1)},t_0+t_{ex},\mathbf{X}[N_{K+1}]^{(K)},t_f)=\boldsymbol{0} &\\
   & \mathbf{X}[i]_{min}^{(k)} \leq \mathbf{X}[i]^{(k)} \leq \mathbf{X}[i]_{max}^{(k)} &\\
   & \mathbf{U}[i]_{min}^{(k)} \leq  \mathbf{U}[i]^{(k)} \leq \mathbf{U}[i]_{max}^{(k)} &\label{eq:boundary_ps}\\
   & t_{f_{min}} \leq  t_f \leq  t_{f_{max}} \label{eq:time_constraint_ps} &
\end{align}
\else
\begin{align}
   & \underset{\mathbf{X}^{(k)},\;{\boldsymbol{U}^{(k)}},\;t_f}{\textbf{minimize}}
   && {\cal{M}}(\mathbf{X}[1]^{(1)},t_0+t_{ex},\mathbf{X}[N_{K+1}]^{(K)},t_K)+ I \label{eq:cost_ps}\\
   & \textbf{subject to}
   && \sum_{j=1}^{N_k+1}\mathbf{X}_{j}^{(k)}D_{ij}^{(k)}-\frac{t_f-t_0+t_{ex}}{2}\mathbf{f}(\mathbf{X}_{i}^{(k)},\mathbf{U}_{i}^{(k)},\tau_i^k;t_0+t_{ex},t_f)=\mathbf{0} \\
   &&& \mathbf{C}^{(k)}(\mathbf{X}[i]^{(k)},\mathbf{U}[i]^{(k)},\tau_{i}^k;t_0+t_{ex},t_f) \le \mathbf{0} \\
   &&& \boldsymbol{\phi}(\mathbf{X}[1]^{(1)},t_0+t_{ex},\mathbf{X}[N_{K+1}]^{(K)},t_f)=\boldsymbol{0}\\
   &&& \mathbf{X}[i]_{min}^{(k)} \leq \mathbf{X}[i]^{(k)} \leq \mathbf{X}[i]_{max}^{(k)} \\
   &&& \mathbf{U}[i]_{min}^{(k)} \leq  \mathbf{U}[i]^{(k)} \leq \mathbf{U}[i]_{max}^{(k)} \label{eq:boundary_ps}\\
   &&& t_{f_{min}} \leq  t_f \leq  t_{f_{max}} \label{eq:time_constraint_ps}
\end{align}
\fi
for $(i = 1,\dots, N_k)$ and $(k = 1,\dots,K)$

%\paragraph{Costate Estimates}
%Estimates of the costates can be utilized to help determine if the resulting solution is optimal \cite{huntington2007advancement}.

%\paragraph{First-Order Optimality}
%relationship between KKT conditions of NLP and first-order optimality conditions of the OCP...

%\subsection{From a Discrete OCP to a Large and Sparse NLP}
\subsubsection{Transforming to an NLP}
Depending on the method, either the discrete OCP in Eqn. \ref{eq:costd_tm} - Eqn. \ref{eq:time_constraintd_tm} or the discrete OCP in  Eqn. \ref{eq:cost_ps} - Eqn. \ref{eq:time_constraint_ps} is then transformed into a large and sparse NLP given by Eqn. \ref{eq:cost_nlp} - Eqn. \ref{eq:eq_nlp}.

Now that design and methods of \NLOpt have been provided, the following two sections compares its ease of use and speed to existing commonly used optimal control software.
\section{Evaluation description} \label{sec:evaluation}
The next section compares \NLOpt and \PROPT in terms of ease of use and speed.
This section describes the conditions under which these comparisons are made.

\subsection{Ease of use}
Claiming that a software package is easy to use is subjective; even with the definition provided for ease of use, i.e., syntax that closely resembles the underlying OCP.
Therefore, the respective syntax in \NLOpt and \PROPT needed to model the moon lander OCP, as given in Eqn. \ref{eq:ml}, is compared.

\subsection{Benchmark}
The conditions under which \NLOptnsp's speed is benchmarked against \PROPT include the benchmark problem, methodology, and setup.

\subsubsection{Benchmark problem}
An OCP suitable for an NMPC-based ground vehicle application is used to benchmark \NLOpt against \PROPTnsp.
The purpose of this problem is to find the steering and acceleration commands that drive a kinematic bicycle model \cite{rajamani2011vehicle,gonzales2016} to a goal location ($x_g=0\;\mathrm{m}$, $y_g=100\;\mathrm{m}$) as fast as possible (i.e., in minimum time) while avoiding crashing into a static obstacle.
The cost functional is shown in Eqn. \ref{eq:kb_cost} as
\begin{equation}
  \begin{aligned}
     & \underset{a_x(t),\;\alpha(t)}{\textbf{minimize}}
     & & (x(t_f)-x_g)^2 + (y(t_f)-y_g)^2 + t_f \label{eq:kb_cost}
  \end{aligned}
\end{equation}

The dynamic constraints are shown in Eqn. \ref{eq:kb_dyn} as
\begin{equation}
  \begin{aligned}
      \dot{x}(t)  &= u_x(t)\cos(\psi(t) + \beta(t)    ) \\
      \dot{y}(t)   &= u_x(t)\sin(\psi(t) + \beta(t) ) \\
      \dot{\psi}(t) &= \frac{u_x(t)\sin(\beta(t))}{l_b} \\
      \dot{u}_x(t)  &= a_x(t)  \label{eq:kb_dyn}
  \end{aligned}
\end{equation}
where $x(t)$ and $y(t)$ are the position coordinates, $\psi(t)$ is the yaw angle, $u_x(t)$ is the longitudinal velocity, $\alpha(t)$ is the steering angle, $\beta(t) = \tan( \frac{l_a \tan(\alpha(t)) }{l_a+l_b} )^{-1} $, $l_a=1.58\;\mathrm{m}$ and $l_b=1.72\;\mathrm{m}$ are the distances from the center of gravity to the front and rear axles, respectively.

The path constraints ensure that the vehicle avoids an obstacle, these constraints are shown in Eqn. \ref{eq:kb_path} as
\begin{equation}
1 < ({\frac{{x(t) - {x}_{obs}}}{{a_{obs} + m}}})^2 + ({\frac{{y(t) - {y}_{obs}}}{{b_{obs} + m}}})^2  \label{eq:kb_path}
\end{equation}
where $x_{obs}=0\;\mathrm{m}$ and $y_{obs}=50\;\mathrm{m}$ denote the position of the center of the obstacle, $a_{obs}=5\;\mathrm{m}$ and $b_{obs}=5\;\mathrm{m}$ denote the semi-major and semi-minor axes, $m=2.5\;\mathrm{m}$ is the safety margin that accounts for the footprint of the vehicle.

The event constraints ensure that the vehicle starts at a particular initial condition, these constraints are given in Eqn. \ref{eq:kb_path} as
\begin{equation}
  \begin{aligned}
  x(t_0) &= 0\;\mathrm{m},  &y(t_0) = 0\;\mathrm{m},   &&\psi(t_0) = \frac{\pi}{2}\;\si{rad} \\
   u_x(t_0) &= 15\;\si{\frac{m}{s}},  &a_x(t_0)= 0\;\si{\frac{m}{s^2}},  &&\alpha(t_0)=0\;\si{rad}  \label{eq:kb_event}
  \end{aligned}
\end{equation}
That is the vehicle is traveling straight ahead at a constant velocity of $15\;\si{\frac{m}{s}}$.

The state and control bound constraints are given in Eqn. \ref{eq:kb_st_ctr} as
\begin{equation}
  \begin{aligned}
      -100\;\mathrm{m}       &\leq x(t) \leq 100\;\mathrm{m},         & -0.01\;\si{m}        \leq y(t) \leq 120\;\mathrm{m}  \\
       -2\pi\;\si{\radian}   & \leq \psi(t) \leq 2\pi\;\si{\radian},  & 5\;\si{\frac{m}{s}}    \leq u_x(t) \leq 29\;\si{\frac{m}{s}} \\
     -2\;\si{\frac{m}{s^2}} & \leq a_x(t)  \leq 2\;\si{\frac{m}{s^2}}, &  \frac{-30\pi}{180}\;\si{\radian}  \leq \alpha(t) \leq \frac{30\pi}{180}\;\si{\radian} \label{eq:kb_st_ctr}
  \end{aligned}
\end{equation}

The final time is constrained to be $ 0.001\;\si{s}\leq t_f \leq 50\;\si{s}$.

Solutions to Eqn. \ref{eq:kb_cost} - Eqn. \ref{eq:kb_st_ctr} that are obtained in less than $0.5\;\si{s}$ are deemed to be fast enough for real-time NMPC.

\subsubsection{Benchmark methodology}
Using the problem described above, a comprehensive benchmark is made between various solvers.
A solver is defined by a particular combination of either \NLOpt or \PROPT in conjunction with a particular direct-collocation method.
The set of solvers $S$ are listed in Table \ref{tab:leg} as

\begin{minipage}{\linewidth}
\begin{footnotesize}
\begin{center}
\captionof{table}{Set of solvers tested} \label{tab:leg}
\vspace{-0.2cm}
\begin{tabular}{ c l }
\hline
Legend label & Description \\
\hline
\nla & \NLOpt with LGR nodes with a single interval \\
\nlb & \NLOpt with LGR nodes with two intervals\\
\nlc & \NLOpt with LGR nodes with four intervals\\
\nleu & \NLOpt using Euler's method \\
\nltr & \NLOpt using trapezoidal method\\
\propta & \PROPT with Chebyshev nodes with a single phase\\
\proptb & \PROPT with Chebyshev nodes with two phases\\
\proptc & \PROPT with Chebyshev nodes with four phases\\
\end{tabular}
\end{center}
\end{footnotesize}
\end{minipage}
\vspace{0.2cm}

Comparisons between the average solve-times of single interval/phase solvers (i.e., \nleunsp, \nltrnsp, \nlansp, \proptansp) and the multiple interval/phase solvers (i.e., \nlbnsp, \nlcnsp, \proptbnsp, \proptc) must be considered in context.
This is true because as the number of collocation points per interval/phase is increased, the two interval/phase solvers (i.e., \nlb and \proptbnsp) and the four interval/phase solvers (i.e., \nlc and \proptcnsp) are solving problems that roughly two and four times larger than the single interval/phase solvers, respectively.
However, there are advantages of these multi interval/phase solvers, as discussed previously, that may be more important than the decreases in solve times.
Thus, these solvers are included in the comparison here for a more comprehensive comparison.
%While such comparisons are not fair, they show how an appropriate solver can be selected for a particular type of problem.
% and (2) to provide trend information for various direct-collocation methods.

Comparisons between the average solve-times of the multiple interval solvers in \NLOpt and the multiple phase solvers in \PROPT also require consideration.
Ideally, the benchmark between \PROPT and \NLOpt would include the same direct-collocation methods.
Unfortunately, \PROPT and \NLOpt do not have the same direct-collocation methods.
As such, comparisons are made between single/multiple phase Chebyshev pseudospectral methods in \PROPT and multiple single/interval LGR pseudospectral methods in \NLOptnsp.
%
%Differences between a multiple phase and multiple interval methods are that, in a multiple phase problem, between phases, the constraints can change and the optimal transition time can be  determined.
Unlike a multiple interval method, in a multiple phase method, between phases, the constraints can change and the optimal transition time can be  determined.
In this work, the constraints do not change and the final time is divided evenly by the number of phases to determine the transition time.
By doing this, the OCPs formulated by the multiple phase and multiple interval methods have roughly the same size and level of complexity.
Thus, comparisons between the two software packages can be made with this issue in mind.

Each solver $s$ is used to solve a set of problems $P$.
The benchmark problem is discretized over the range of collocation points $p=2,3,\hdots,102$ per interval or phase to realize the set of problems $P$ tested for each solver; a total of $101$ different values of $p$ (i.e., levels-of-fidelity or problems) are tested.
Each test is performed three times to provide the data needed to calculate the average solve-time $t_{s,p}$ for the benchmark problem with a level-of-fidelity $p$ using solver $s$.
A polynomial is interpolated through the $(x,y)$ solution points and sampled at $200$ points to determine if the solution drives the vehicle through the obstacle.
If a collision is determined for a particular combination of solver $s^*$ and level-of-fidelity $p^*$, then $t_{s^*,p^*}$ is set to $NaN$; such solutions are not practically feasible.

Conducting many benchmark tests helps accurately rank the solvers.
However, analyzing large sets of benchmark data can be overwhelming and the conclusions drawn from such analyses can be subjective.
To help eliminate these issues, this work uses an optimization software benchmarking tool called performance profiles \cite{dolan2001}.

%A performance profile shows the variation of a particular performance metric.
%Performance profiles analyze benchmark data to show the trade-off
Performance profiles show the distribution function for a particular performance metric.
Here, the performance metric is the ratio of the solver's average solve-time to the best average solver solve-time given as
\begin{equation*}
r_{s,p} = \frac{t_{s,p}}{\text{min}(t_{s,p}:s \in S)}
\end{equation*}
where this performance metric is calculated for each solver $s$ at each level-of-fidelity $p=2,3,\hdots,102$.
If a solver does not solve a particular problem, then $r_{s,p}$ is set to $r_M$. $r_M$ is chosen to be a large positive number; the choice of $r_M$ does not effect the evaluation \cite{dolan2001}.

To assess a solver's overall performance on the set of problems, the cumulative distribution function for the performance ratio is defined as
\begin{equation*}
\mathcal{P}_s(\Gamma) = \frac{1}{101}{\text{size}(p \in P: r_{s,p} \leq \Gamma)}
\end{equation*}
where $\mathcal{P}_s(\Gamma)$ is the probability that solver $s$ can solve problem $p$ within a factor $\Gamma$ of the best ratio.
%\begin{equation}
%  P(r_{p,s} \leq \Gamma :1 \leq s \leq n_s)
%\end{equation}
%Performance profiles analyze benchmark data to allows a performance metric to be solvers to visually compares the solvers.
%For this data, performance profiles are used to visualize the trade-off between speed and reliability.
%For a given benchmark problem, a particular combination of either \NLOpt or \PROPTnsp, direct-collocation method, and number of collocation points
%In this work, reliability is defined as the probability that a given problem type will be solved the best (e.g., fastest) for a particular
%. Problem type is a given OCP approximated using different levels of fidelity (i.e., collocation points).
%A methodology is then provided to visualize the trade-off between speed and reliability that is established by the combinations of software, direct-collocation methods
\subsubsection{Setup}
%\paragraph{Hardware Platform and Software Stack}
The setup is defined by the hardware platform and software stack. The results in this paper are produced using a single machine running Ubuntu $16.04$ with the following hardware characteristics; an Intel Core $\mathrm{i}7-4910\mathrm{MQ}$ CPU $@ 2.90\mathrm{GHz} \times 8$, and $16\mathrm{GB}$ of $\mathrm{RAM}$. For software, both \NLOpt $0.1.5$ and \PROPT use \KNITRO $10.3$ for the NLP solver with the default settings, except the maximum solve-time, which is set to $300\;\si{s}$.
\section{Results} \label{sec:results}

\subsection{Ease of use}
Listing. \ref{lst:nl_moon} and Listing. \ref{lst:pr_moon} show the respective syntax in \NLOpt and \PROPT needed to model the moon lander OCP in Eqn. \ref{eq:ml}. Section~\ref{sec:moon} in the Appendices shows \NLOptnsp's and \PROPTnsp's solutions compared to the analytical solution.
\if\slides1
\begin{minipage}{\linewidth}
\begin{lstlisting}[caption={\NLOpt code needed to formulate and solve the moon lander problem.},label=lst:nl_moon,frame=lines]
n = define(numStates = 2, numControls = 1, X0 = [10, -2], XF = [0., 0.], XL = [0, -20], XU = [20, 20], CL = [0.], CU=[3.]);
dynamics!(n, [:(x2[j]), :(u1[j] - 1.5)]);
configure!(n; (:finalTimeDV => true));
obj = integrate!(n, :(u1[j]));
@NLobjective(n.ocp.mdl, Min, obj);
optimize!(n);
\end{lstlisting}
\end{minipage}
\else
%\begin{figure}
\begin{lstlisting}[caption={\NLOpt code needed to formulate and solve the moon lander problem. The ! character indicates that the function is modifying the model.},label=lst:nl_moon,frame=lines]
n = define(numStates = 2, numControls = 1, X0 = [10, -2], XF = [0., 0.], XL = [0, -20], XU = [20, 20], CL = [0.], CU=[3.]);
dynamics!(n,[:(x2[j]), :(u1[j] - 1.5)]);
configure!(n;(:finalTimeDV => true));
obj = integrate!(n, :(u1[j]));
@NLobjective(n.ocp.mdl, Min, obj);
optimize!(n);
\end{lstlisting}
%\caption{The ! character on the functions means that the model is being modified.}
%\end{figure}
\fi

%dx = [:(x2[j]),:(u1[j]-1.5)]
% where, by default, when the final time is a design variable it is constrained to be between $0.001$ and $1000$

\if\slides1
\begin{minipage}{\linewidth}
\begin{lstlisting}[caption={\PROPT code needed to formulate and solve the moon lander problem},label=lst:pr_moon,frame=lines]
toms t t_f
p = tomPhase('p', t, 0, t_f, 30);
setPhase(p);
tomStates x v
tomControls a
cbox = {0.001<=t_f<=400, 0<=icollocate(x)<=20, -20<=icollocate(v)<= 20, 0<=collocate(a)<=3};
ode = collocate({dot(x)==v, dot(v)==-1.5 + a});
cbnd = {initial(x == 10);initial(v == -2);final(x == 0);final(v == 0);};
x0 = {t_f == 1.5, icollocate({x == 0 v == 0}), collocate(a == 0)};
objective = integrate(a);
options = struct;
prob = sym2prob(objective, {cbox, ode, cbnd}, x0, options);
result = tomRun('knitro', prob, 1);
\end{lstlisting}
\end{minipage}
\else
\begin{figure*}
\begin{lstlisting}[caption={\PROPT code needed to formulate and solve the moon lander problem},label=lst:pr_moon,frame=lines]
toms t t_f
p = tomPhase('p', t, 0, t_f, 30);
setPhase(p);
tomStates x v
tomControls a
cbox = {0.001<=t_f<=400, 0<=icollocate(x)<=20, -20<=icollocate(v)<= 20, 0<=collocate(a)<=3};
ode = collocate({dot(x)==v, dot(v)==-1.5 + a});
cbnd = {initial(x == 10);initial(v == -2);final(x == 0);final(v == 0);};
x0 = {t_f == 1.5, icollocate({x == 0 v == 0}), collocate(a == 0)};
objective = integrate(a);
options = struct;
prob = sym2prob(objective, {cbox, ode, cbnd}, x0, options);
result = tomRun('knitro', prob, 1);
\end{lstlisting}
\end{figure*}
\fi
%options.solver= 'knitro';
%options.use_d2c = true;
% solution = getSolution(result);

\if\showoutline1
\begin{outline}[enumerate]
\1 \NLOpt can model OCPs more succinctly than \PROPTnsp.
  \2 \NLOpt models Eqn. \ref{eq:ml} with $5$ lines of code, while it takes \PROPT $12$ lines --- there are two main reasons for this:
    \3 (1) it takes \PROPT $4$ lines of code to include the initial and final state conditions, and the upper and lower limits of the states and controls, while this is accomplished with a single line of code using \NLOptnsp,
    \3 and (2) several of \PROPTnsp's features are required, while in \NLOpt they are optional: the initial guess needs to be passed, an options structure needs to be passed, and the state and control variables need to be given a particular name.
  \2 Additionally, \PROPT has more verbose syntax than \NLOptnsp.
    \3 \PROPTnsp's \lstinline!collocate()!, \lstinline!initial()!, and \lstinline!final()! functions require many characters per line of code.
\end{outline}
\else
\NLOpt can model OCPs more succinctly than \PROPTnsp. \NLOpt models Eqn. \ref{eq:ml} with $5$ lines of code, while it takes \PROPT $12$ lines --- there are two main reasons for this: (1) it takes \PROPT $4$ lines of code to include the initial and final state conditions, and the upper and lower limits of the states and controls, while this is accomplished with a single line of code, with \NLOptnsp, and (2) several of \PROPTnsp's features are required, while in \NLOpt they are optional; these features include an initial guess, an options structure, and the naming of the state and control variables. Additionally, \PROPT has more verbose syntax than \NLOpt --- \PROPTnsp's \lstinline!collocate()!, \lstinline!initial()!, and \lstinline!final()! functions require many characters per line of code.
\fi

\subsection{Speed}
The performances of the solvers in Table \ref{tab:leg} are now examined on the set of problems realized by various discretizations of Eqn. \ref{eq:kb_cost} -- Eqn. \ref{eq:kb_st_ctr}, as described in Sec. \ref{sec:evaluation}.
The results for these examinations are in Fig. \ref{fig:kbe} and Fig. \ref{fig:kbf}.
Fig. \ref{fig:kbe} shows the performance, or average solve-times $t_{s,p}$, for each solver $s$ on each problem $p$.
Fig. \ref{fig:kbf} shows the performance profiles for all of the solvers in four ranges of interest for $\Gamma$. Each range is on a separate plot.
The purpose of this section is to (1) show the raw benchmark data in Fig. \ref{fig:kbe} and (2) provide an objective analysis of this data in Fig. \ref{fig:kbf}.
In the following section, this information will be used to draw conclusions regarding the speed of \NLOpt and the best solver for the benchmark problem.
%suitability of the performance profiles to help determine an appropriate solver for a particular type of NMPC problem.
% D: another method may be better with a particular number of collocation points
% emphasize that this is a single problem...
%In the description of the results that follows,
%that the this work is not claiming that \NLOpt can solve all NMPC problems faster than \PROPTnsp.
% On the contrary,

% D The benchmark indicates that \NLOpt is fast; overall,
Fig. \ref{fig:kbe} shows that \NLOptnsp's solvers are faster than \PROPTnsp's.
At a high-level, \NLOpt solves $88\%$ of the problems in real-time using h-methods (i.e., \nleu and \nltrnsp) and $46\%$ of the time using p$/$hp-methods (i.e., \nlansp, \nlbnsp, and \nlcnsp).
\PROPT only solves $0.05\%$ of the problems in real-time using p$/$hp-methods (i.e., \proptansp, \proptbnsp, and \proptcnsp).
At a lower-level, the zoomed-in subplot in the the bottom graph of Fig. \ref{fig:kbe} shows that \NLOpt solves the benchmark problem in real-time when the number of collocation points per interval is less than: $80$ for the single-interval case; $45$ for the two-interval case; and $25$ for the four-interval interval case.
\PROPT obtains real-time solutions when the number of collocation points per phase is less than $27$ for the single-phase case and less than $4$ for the two-phase case.
For the four-phase case, \PROPT cannot solve any of the problems in real-time.

Fig. \ref{fig:kbe} also shows that as the number of intervals/phases increase from \nla to \nlc and \propta to \proptcnsp, the solve-times increase exponentially.
Due to the large solve-times with \PROPT's solvers, these trends can only be seen in the top graph of Fig. \ref{fig:kbe} --- the bottom graph shows the trends for \NLOptnsp's solvers.
As discussed in the previous section, this increase in solve time is largely due to the fact that with an increase in the intervals/phases larger problems are created and they take longer to solve.
Even though the \nlc solver is solving a problem that is roughly four times larger than the \propta solver, \nlc results in smaller solve-times.

Fig. \ref{fig:kbe} also shows that h-methods in \NLOpt are faster than the p-method for the benchmark problem.
As the level-of-fidelity increases, the solve-times increase linearly with h-methods and exponentially with the p-method.
Additionally, the number of collocation points needs to be greater than about $20$ for the h-methods to ensure collision avoidance, while the p-methods need $23$ and $27$ for \NLOpt and \PROPTnsp, respectively.
%
%The \proptc solver is the only solver that is able to avoid collisions with the obstacle in all cases.
%

The four plots in Fig \ref{fig:kbf} show the ranges of $\Gamma$ wherein certain solvers dominate.
Each profile in this figure shows the probability $\mathcal{P}$ that a given solver $s$ will solve the set of problems $P$ the fastest within a factor of $\Gamma$.
At $\Gamma=1$, the solver that has the highest probability of being the fastest is \nleunsp, with a probability of $0.881$.
\nleu dominates until about $\Gamma=1.8$, at which point \nltr has the highest probability of being the fastest, with a probability of $0.891$.
\nltr dominates until about $\Gamma=80$.
The remaining approximate ranges of domination are as follows: \nlb from $80$ to $160$, \nlc from $160$ to $5,000$, \proptc from $5,000$ onwards.
%
%The \proptc solver is the only solver that is able to avoid collisions with the obstacle in all cases.
Given enough time, \proptc solves $100\%$ of the problems.
For this benchmark problem, while the \nltr and \nleu solvers are much faster than the \nlbnsp, \nlcnsp, and \proptc solvers they are not as reliable.
However, the \nltr and \nleu solvers are both faster and more reliable than the \nlansp, \proptansp, and \proptb solvers.

% read in data and create macros Bench
\pgfplotstableread{Bicycle/benchmark1/results/PROPT_single.csv}{\prsingle}
\pgfplotstableread{Bicycle/benchmark1/results/PROPT_two.csv}{\prtwo}
\pgfplotstableread{Bicycle/benchmark1/results/PROPT_four.csv}{\prfour}

\pgfplotstableread{Bicycle/benchmark1/results/NL_lgrExplicit_1.csv}{\nlsingle}
\pgfplotstableread{Bicycle/benchmark1/results/NL_lgrExplicit_2.csv}{\nltwo}
\pgfplotstableread{Bicycle/benchmark1/results/NL_lgrExplicit_4.csv}{\nlfour}
\pgfplotstableread{Bicycle/benchmark1/results/NL_trapezoidal.csv}{\nlt}
\pgfplotstableread{Bicycle/benchmark1/results/NL_bkwEuler.csv}{\nle}

% read in data and create macros PP
\pgfplotstableread{Bicycle/benchmark1/results/PP_5.dat}{\ppa}
\pgfplotstableread{Bicycle/benchmark1/results/PP_50.dat}{\ppb}
\pgfplotstableread{Bicycle/benchmark1/results/PP_200.dat}{\ppc}
\pgfplotstableread{Bicycle/benchmark1/results/PP_8000.dat}{\ppd}

\if\slides1
% bench
\newcommand{\hbench}{4cm}
\newcommand{\wbench}{10cm}

% pp
\newcommand{\hpp}{4cm}
\newcommand{\wpp}{6.5cm}
\newcommand{\ys}{-0.5ex}
\newcommand{\hsa}{0.6cm}
\newcommand{\vsa}{0.6cm}

\newcommand{\yleg}{$P$}
\else
  \if\doublecol1
    % bench
    \newcommand{\hbencha}{4cm}
    \newcommand{\hbenchb}{4cm}
    \newcommand{\wbench}{7cm}

    % pp
    \newcommand{\hpp}{4.5cm}
    \newcommand{\wpp}{4.5cm}
    \newcommand{\ys}{1ex}
    \newcommand{\hsa}{0.6cm}
    \newcommand{\vsa}{0.6cm}
  \else
    % bench
    \newcommand{\hbencha}{3cm}
    \newcommand{\hbenchb}{5cm}
    \newcommand{\wbench}{8.5cm}

    % pp
    \newcommand{\hpp}{5cm}
    \newcommand{\wpp}{5cm}
    \newcommand{\ys}{1ex}
    \newcommand{\hsa}{0.6cm}
    \newcommand{\vsa}{0.6cm}
  \fi
\newcommand{\yleg}{$\mathcal{P}$}

\fi

\begin{figure}[!hpb]
\centering
% https://tex.stackexchange.com/questions/103060/left-margin-minipage-two-figures
\begin{minipage}[b]{0.9\linewidth}
\centering

\begin{subfigure}[t]{\textwidth}
\begin{tikzpicture}[only marks]

\begin{groupplot}[
group style={group name = myplot, group size = 1 by 2, horizontal sep = 0.4cm, vertical sep = 4pt},
xmin = 2,
xmax = 102,
ymin = 0,
grid=major,
cycle list name = my markers,
ylabel={Solve-time $(\si{s})$}]

\nextgroupplot[
ymax = 300,
xticklabels=\empty,
height = \hbencha,
width = \wbench]
\addplot +[thin] table [x=Nck, y=t] {\nlsingle}; \label{plots:bbia}
\addplot +[thin] table [x=Nck, y=t] {\nltwo};\label{plots:bbib}
\addplot +[thin] table [x=Nck, y=t] {\nlfour};\label{plots:bbic}
\addplot +[thin] table [x=Nck, y=t] {\nlt};\label{plots:bbid}
\addplot +[thin] table [x=Nck, y=t] {\nle};\label{plots:bbie}
\addplot +[thin] table [x=Nck, y=t] {\prsingle};\label{plots:bbif}
\addplot +[thin] table [x=Nck, y=t] {\prtwo};\label{plots:bbig}
\addplot +[thin] table [x=Nck, y=t] {\prfour};\label{plots:bbih}
\coordinate (top) at (rel axis cs:0,1);% coordinate at top of the first plot

\nextgroupplot[ymax = 0.7,
            xlabel={Number of collocation points per interval/phase},
            height = \hbenchb,
            width = \wbench
             ]
\addplot +[thin,y filter/.expression={y>0.7 ? NaN : y}] table [x=Nck, y=t] {\nlsingle};
\addplot +[thin,y filter/.expression={y>0.7 ? NaN : y}] table [x=Nck, y=t] {\nltwo};
\addplot +[thin,y filter/.expression={y>0.7 ? NaN : y}] table [x=Nck, y=t] {\nlfour};
\addplot +[thin,y filter/.expression={y>0.7 ? NaN : y}] table [x=Nck, y=t] {\nlt};
\addplot +[thin,y filter/.expression={y>0.7 ? NaN : y}] table [x=Nck, y=t] {\nle};
\addplot +[thin,y filter/.expression={y>0.7 ? NaN : y}] table [x=Nck, y=t] {\prsingle};
\addplot +[thin,y filter/.expression={y>0.7 ? NaN : y}] table [x=Nck, y=t] {\prtwo};
\addplot +[thin,y filter/.expression={y>0.7 ? NaN : y}] table [x=Nck, y=t] {\prfour};
\coordinate (bot) at (rel axis cs:1,0);% coordinate at bottom of the last plot

\end{groupplot}
% legend
\path (top|-current bounding box.north)--
      coordinate(legendpos)
      (bot|-current bounding box.north);
\matrix[
    matrix of nodes,
    anchor=south,
    draw,
    inner sep=0.2em,
    draw,
  ]at([xshift=0cm,yshift=0cm]legendpos)
  {
    \ref{plots:bbia}& \getdata[1]\NLstringsB &[5pt]
    \ref{plots:bbib}& \getdata[2]\NLstringsB &[5pt]
      \ref{plots:bbic}& \getdata[3]\NLstringsB &[5pt] \\
    \ref{plots:bbid}& \getdata[4]\NLstringsB &[5pt]
      \ref{plots:bbie}& \getdata[5]\NLstringsB &[5pt]
    \ref{plots:bbif}& \getdata[1]\PRstringsB &[5pt] \\
      \ref{plots:bbig}& \getdata[2]\PRstringsB &[5pt]
    \ref{plots:bbih}& \getdata[3]\PRstringsB &[5pt]
                     &                        &     \\};

\end{tikzpicture}

\if\slides0
	\caption{The top graph shows all average solve-times obtained. The bottom graph shows the solve-times obtained near or below the real-time threshold of $0.5\;\si{s}$. \label{fig:kbe}}
\fi
\end{subfigure}
\begin{subfigure}[t]{\textwidth}
\begin{tikzpicture}[]
\begin{groupplot}[group style={group name = myplot, group size = 2 by 2, horizontal sep = \hsa, vertical sep = \hsa}, height = \hpp, width = \wpp]
\nextgroupplot[
             xmin = 1,
             xmax = 5,
             ymin = 0,
             ymax = 1,
   		     ylabel=\yleg,
    	     grid=major,
             cycle list name=my markers
             ]

\addplot +[thin] table [x=tau, y=nla] {\ppa};\label{plots:bif}
\addplot +[thin] table [x=tau, y=nlb] {\ppa};\label{plots:big}
\addplot +[thin] table [x=tau, y=nlc] {\ppa};\label{plots:bih}
\addplot +[thin] table [x=tau, y=nlt] {\ppa};\label{plots:bii}
\addplot +[thin] table [x=tau, y=nle] {\ppa};\label{plots:bij}
\addplot +[thin] table [x=tau, y=pra] {\ppa};\label{plots:bik}
\addplot +[thin] table [x=tau, y=prb] {\ppa};\label{plots:bil}
\addplot +[thin] table [x=tau, y=prc] {\ppa};\label{plots:bim}

\coordinate (top) at (rel axis cs:0,1);% coordinate at top of the first plot

\nextgroupplot[
             xmin = 1,
             xmax = 50,
             ymin = 0,
             ymax = 1,
    	     grid=major,
             cycle list name=my markers,
             ]
\addplot +[thin] table [x=tau, y=nla] {\ppb};
\addplot +[thin] table [x=tau, y=nlb] {\ppb};
\addplot +[thin] table [x=tau, y=nlc] {\ppb};
\addplot +[thin] table [x=tau, y=nlt] {\ppb};
\addplot +[thin] table [x=tau, y=nle] {\ppb};
\addplot +[thin] table [x=tau, y=pra] {\ppb};
\addplot +[thin] table [x=tau, y=prb] {\ppb};
\addplot +[thin] table [x=tau, y=prc] {\ppb};

\nextgroupplot[
             xmin = 1,
             xmax = 200,
             ymin = 0,
             ymax = 1,
             xlabel = {$\Gamma$},
   		     ylabel=\yleg,
    	     grid=major,
             cycle list name=my markers,
             ]
\addplot +[thin] table [x=tau, y=nla] {\ppc};
\addplot +[thin] table [x=tau, y=nlb] {\ppc};
\addplot +[thin] table [x=tau, y=nlc] {\ppc};
\addplot +[thin] table [x=tau, y=nlt] {\ppc};
\addplot +[thin] table [x=tau, y=nle] {\ppc};
\addplot +[thin] table [x=tau, y=pra] {\ppc};
\addplot +[thin] table [x=tau, y=prb] {\ppc};
\addplot +[thin] table [x=tau, y=prc] {\ppc};

\nextgroupplot[
             xtick = {0,4000,8000},
             xmin = 1,
             xmax = 8000,
             ymin = 0,
             ymax = 1,
             xlabel = {$\Gamma$},
    	     grid=major,
             cycle list name=my markers,
             ]
\addplot +[thin] table [x=tau, y=nla] {\ppd};
\addplot +[thin] table [x=tau, y=nlb] {\ppd};
\addplot +[thin] table [x=tau, y=nlc] {\ppd};
\addplot +[thin] table [x=tau, y=nlt] {\ppd};
\addplot +[thin] table [x=tau, y=nle] {\ppd};
\addplot +[thin] table [x=tau, y=pra] {\ppd};
\addplot +[thin] table [x=tau, y=prb] {\ppd};
\addplot +[thin] table [x=tau, y=prc] {\ppd};
\coordinate (bot) at (rel axis cs:1,0);% coordinate at bottom of the last plot

\end{groupplot}

\end{tikzpicture}
\if\slides0
	\caption{Performance profiles: each graph shows the performance profiles for a different range of $\Gamma$.\label{fig:kbf}}
\fi
\end{subfigure}
\if\slides0
	\caption{Benchmark results \NLOpt and \PROPT for the kinematic bicycle problem, see Table \ref{tab:leg} for legend explanation.}
\fi
\end{minipage}

\end{figure}
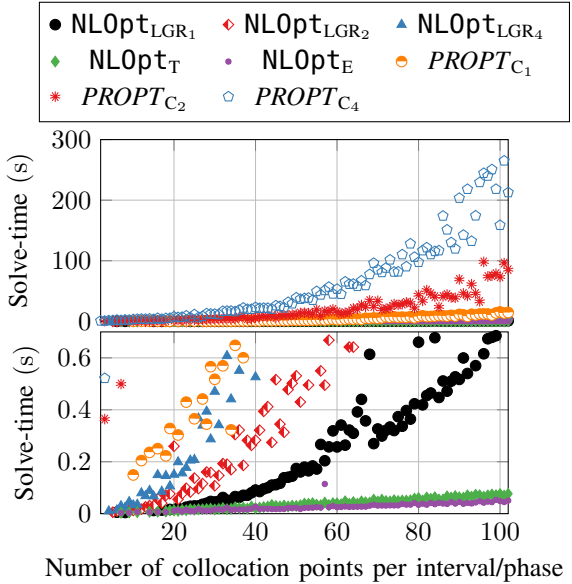
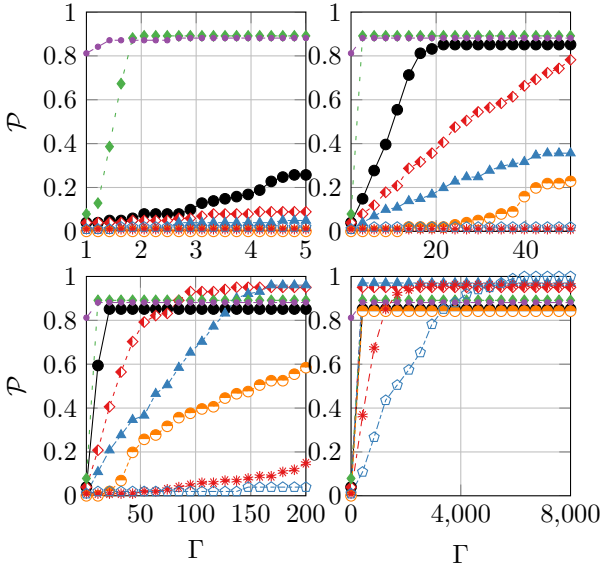
% https://tex.stackexchange.com/questions/78098/captioning-nested-subfigures-with-subcaption

%First, we evaluate now NLOptControl's ability to translate an optimal control model suitable for an NMPC application into an NLP solver format that can be solved: (1) quickly enough for real-time, (2) faster than \PROPTnsp, a state-of-the-art commercial direct-collocation-based optimal control software package, and (3)
\section{Discussion} \label{sec:discussion}
%%%%%%%%%%%%%%%%
The approach detailed in Section \ref{sec:approach} yields a direct-collocation-based optimal control modeling language that is both faster and easier to use than \PROPTnsp.
The results and the following discussion support this claim.
%This claim is supported by the results
%, which demonstrate that \NLOpt is easier to use and faster than \PROPTnsp.

%%%%%%%%%%%%%%%%%%%
\NLOpt is easier to use than \PROPTnsp, because its syntax is more concise, and focused on building a model of the OCP in Bolza form.
Differences between Listing. \ref{lst:nl_moon} and Listing. \ref{lst:pr_moon}, in terms of number of lines of code and the number of characters per line of code, indicate that \NLOpt models OCPs more succinctly than \PROPTnsp.
%
%While it is not shown in this paper, it takes $20$ and $50$ lines of code to formulate the kinematic bicycle problem, using \PROPT and \NLOpt, respectively.
%
This work speculates that \PROPT requires more lines of code to formulate other more practical problems as well.
%

%%%%%%%%%%%%%%%%%%%%%%%%%%
In addition to \PROPTnsp's verbosity, its syntax is flexible to the extent that modeling errors are easier to be made.
This claim is made because its users can more easily formulate problems that do not fit into the Bolza OCP form.
As an example, consider using \PROPT to model the dynamic constraints in Eqn. \ref{eqn:dynamics} for the moon lander problem --- Line $7$ in Listing. \ref{lst:pr_moon}. When using \PROPTnsp, if the user were to forget to include the second differential equation as
\begin{lstlisting}
ode = collocate({dot(x)==v});
\end{lstlisting}
an error would not be displayed; such overly flexible syntax can lead to modeling errors. If that same mistake were attempted in \NLOptnsp, the user would be alerted as
\begin{lstlisting}
julia> dynamics!(n,[:(x2[j])]);
ERROR: The number of differential equations must equal ocp.state.num.
\end{lstlisting}
%This work speculates that \NLOpt helps avoid modeling errors better than \PROPT because \NLOpt focuses on modeling problems in the Bolza form, while \PROPT does not.
%\NLOpt helps avoid modeling errors better than \PROPT because \NLOptnsp's syntax does not allow users to formulate problems that are not in the Bolza form, while \PROPTnsp's syntax does.
Thus, \NLOpt helps avoid modeling errors better than \PROPTnsp, because \NLOptnsp's syntax does not allow users to formulate problems that are not in the Bolza form, while \PROPTnsp's syntax does.

%%%%%%%%%%%%%%%%%%%%%%%
%\NLOpt is a modeling language and \PROPT is more of a solver.
%While like \NLOptnsp, \PROPT can formulate OCPs, it takes a functional approach to this task --- rather than a modeling approach.
Both \NLOpt and \PROPT can be used formulate OCPs, but \PROPT takes a functional approach to this task rather than a modeling approach, as with \NLOptnsp.
Listing. \ref{lst:nl_moon} is compared to Listing. \ref{lst:pr_moon} to support this claim.
Listing. \ref{lst:pr_moon} shows that, with \PROPTnsp, the user creates all of the components of the OCP and finally assembles them on Line $12$.
%
%This design can lead to additional modeling errors; especially if the user tries to model and solve multiple problems in the same script --- which may the case in hierarchical model predictive control applications.
%
With \NLOptnsp, in Listing. \ref{lst:nl_moon}, it is clear from the first line of code that a model named \lstinline!n! is being built.
Using this approach, \NLOpt can clearly model and solve multiple OCPs at once.
%
%For instance, changing \lstinline!n! in Listing. \ref{lst:nl_moon} to another model object name, for instance \lstinline!m!, would allow the the user to model and solve both the Bryson-Denham and moon lander problems together.
%While formulating multiple problems at once is possible with \PROPTnsp, the user would have to be particularly careful not to mix up the components of each problem.
Such an object-oriented approach can further reduce potential modeling errors.

%%%%%%%%%%%%%%%%%%%%%%
The benchmark results in Fig. \ref{fig:kbe} and Fig. \ref{fig:kbf} show that \NLOpt is faster than \PROPTnsp.
Differences between these packages that affect speed include: differentiation methods, underlying computational language, and available direct-collocation methods.

\PROPT uses symbolic automatic differentiation to calculate the derivatives.
However, the structure of the Hessian matrices born from approximating an OCP using direct-collocation methods is sparse and symbolic automatic differentiation does not exploit this structure for speed.
In contrast, \NLOpt uses the acyclic-coloring method to exploit the sparse structure of the Hessian matrix in conjunction with reverse automatic differentiation.
Based on this difference, \NLOpt is expected to be faster than \PROPTnsp, especially when solving large problems that have a very sparse structure.

\PROPTnsp's differentiation methods are implemented in \MATLAB and \NLOptnsp's are implemented in \Juliansp.
Unfortunately, the literature does not contain benchmarks of each of these differentiation methods in both \MATLAB and \Juliansp.
However, research has shown that \Julia is much faster than \MATLAB for a wide range of problem types \cite{bezanson2012julia}.
Thus, \Julia may be able to run the reverse automatic differentiation method combined with the acyclic-coloring method to identify sparsity in the Hessian matrix faster than \MATLAB---if it were implemented in \MATLABnsp.

%%%%%%%%%%%
Overall, this paper speculates that \NLOptnsp's unique combination of differentiation methods and computational language makes it faster than \PROPTnsp.
This only a speculation since the direct-collocation methods are different between \NLOpt and \PROPTnsp.
However, the following pairs of solvers can be considered roughly equivalent in terms of their direct-collocation methods: \nla and \proptansp, \nlb and \proptbnsp, and \nlc and \proptcnsp.
Between these pairs, \NLOpt solves the problem roughly $14$, $26$, and $36$ times faster than \PROPTnsp, respectively.
It is unlikely that these large differences are due to either differences between collocating at Chebyshev nodes vs. LGR nodes, multiple interval vs. multiple phase methods, or some combination of the two.
Thus \NLOpt is fast, which is especially important for MPC applications.
A brief discussion of \NLOptnsp's salient MPC functionality follows.

%%%%%%%%%%%%
\NLOpt has optional functionality that helps account for the non-negligible solve-times in MPC applications.
In MPC, often the control delay (i.e., solve-time) is neglected \cite{liu2017combined}.
When the control delay is neglected, the current state of the plant is used to initialize the problem as opposed to initializing the problem with a prediction of what the plant's state will be after the solve-time has elapsed.
Typically neglecting the solve-time in linear model predictive control is not an issue; because in linear MPC, the quadratic program is solved so quickly that initial state of the trajectory can be set to the current state of the plant without compromising robustness \cite{diehl2009efficient}.
However, neglecting the solve-time in NMPC is likely to deteriorate robustness \cite{simon2009swelling}, because the NLPs often take a non-negligible amount of time to solve, after which the state of the plant will have evolved significantly.
To make matters even more challenging, ensuring that the NLP solve-times are smaller than a particular execution horizon remains an unsolved problem \cite{morari1999model,eskandarian2012handbook, ohtsuka2009practical, diehl2009efficient,albersmeyer2009fast,kirches2010efficient}.
Fortunately, for many problems, these NLP solve-times are similar and an upper limit determined based on experience.
This upper limit can be used to determine a fixed execution horizon.
In \NLOptnsp, after the user selects an execution horizon, as described in Section. \ref{sec:scope}, the framework in Fig. \ref{fig:NB} can be used to account for non-negligible solve times.
Frameworks such as this, can help establish conceptual schemes to improve safety and performance in NMPC applications.
\section{Conclusions} \label{sec:conclusion}
This paper introduces an open-source, direct-collocation method based OCP modeling language called \NLOptnsp.
\NLOpt extends the \JuMP optimization modeling language to include a natural algebraic syntax for modeling OCPs.
\NLOpt is compared against \PROPT in terms of ease of use and speed.
%
%The syntax is used to model the Bryson-Denham and moon lander problems.
%
\PROPTnsp's syntax is shown to be more verbose and error-prone than \NLOptnsp's; thus \NLOpt is easier to use than \PROPTnsp.
This ease of use is largely attributed to \NLOptnsp's use of the \JuMP optimization modeling language.
In addition to being easier to use, results from the benchmark tests show that \NLOpt is much faster than \PROPTnsp.
%thus \NLOpt is deemed fast.
\NLOptnsp's superior performance is likely due to the unique utility of the \Julia programming language and the reverse automatic differentiation method in conjunction with the acyclic-coloring method to exploit the sparsity of the Hessian matrices.
%
%The benchmark also investigates the utility of performance profiles in conjunction with raw benchmark data as a method for selecting an appropriate transcription method for a particular type of problem NMPC problem.
%
%The results show that this type of benchmark method is useful --- for instance, for the kinematic ground vehicle application, the benchmark method shows that the trapezoidal method is preferable, given the need for speed and reliability.
%
\NLOpt emerges as an easy to use, fast, and open-source \cite{febbo_2017} optimal control modeling language that holds great potential for not only improving existing off-line and on-line control systems but also engendering a wide variety of new ones.
%, but also engendering a wide variety of off- and on-line control systems.

\section*{Acknowledgements}
This work was supported by the Automotive Research Center (ARC) in accordance with Cooperative Agreement W56HZV-14-2-0001 U.S. Army Tank Automotive Research, Development and Engineering Center (TARDEC) Warren, MI. Beno{\^i}t Legat provided a thoughtful review of this paper and discussions with the several \Julia developers including Miles Lubin, Tony Kelman, and Chris Rackauckas were helpful.

\bibliography{refs}

\if\thesis0
 \begin{appendices}
  \section{Bryson-Denham problem} \label{sec:bryson}
\else
  \appendix{Supplementary Material for Chapter \ref{nlopt}}
\fi
Fig. \ref{fig:bda} shows the analytic solutions for the states, control, as well as costates compared to the results obtained with \NLOpt using a single interval with $30$ LGR nodes.
\NLOpt calculates these trajectories reasonably well.
% read in data and creatcat st_c	e macros
\pgfplotstableread{BrysonDenham/case1/results/sparse_optimal_solution.csv}{\optimal}
\pgfplotstableread{BrysonDenham/case1/results/st_ctr_poly.csv}{\nlopt}
\pgfplotstableread{BrysonDenham/case1/results/st_ctr.csv}{\nloptcp}
%https://tex.stackexchange.com/questions/68577/compiling-a-document-with-pgfplots-processing-only-every-x-th-data-point

\if\slides1
\newcommand{\ha}{3.3cm}
\newcommand{\wa}{6cm}

\newcommand{\hb}{3.7cm}
\newcommand{\wb}{6cm}
\else
  \if\doublecol1
    \renewcommand{\ha}{5cm}
    \renewcommand{\wa}{4.8cm}

    \renewcommand{\hb}{6.1cm}
    \renewcommand{\wb}{4.8cm}

    \newcommand{\hsp}{0.4cm}
    \newcommand{\vsp}{0.4cm}
  \else

    \renewcommand{\ha}{5cm}
    \renewcommand{\wa}{6cm}

    \renewcommand{\hb}{6.1cm}
    \renewcommand{\wb}{6cm}

    \newcommand{\hsp}{0.4cm}
    \newcommand{\vsp}{0.4cm}
  \fi
\fi

\begin{figure}[!htb]

\begin{minipage}[b]{0.5\linewidth}
\centering
\begin{tikzpicture}
\begin{groupplot}[group style={group name = myplot, group size = 1 by 3, horizontal sep = \hsp, vertical sep = \vsp}, height = \ha, width = \wa]
\nextgroupplot[
             xmin = 0,
             xmax = 1,
   		     ylabel={$x(t)$},
   		     xticklabels=\empty,
    	     grid=major,
             cycle list name=my markers % #"mark list*" #"black white"
             ]
\addplot +[mark=none] table [x=t, y=x] {\optimal}; \label{plots:bda}
\addplot +[mark=none] table [x=t, y=x] {\nlopt}; \label{plots:bdb}
\addplot +[only marks,thin] table [x=t, y=x] {\nloptcp}; \label{plots:bdc}

\coordinate (top) at (rel axis cs:0,1);% coordinate at top of the first plot

\nextgroupplot[
             xmin = 0,
             xmax = 1,
   		     ylabel={$v(t)$},
   		     xticklabels=\empty,
    	     grid=major,
             cycle list name=my markers
             ]
\addplot +[mark=none] table [x=t, y=v] {\optimal};
\addplot +[mark=none] table [x=t, y=v] {\nlopt};
\addplot +[only marks,thin] table [x=t, y=v] {\nloptcp};

\nextgroupplot[
             xmin = 0,
             xmax = 1,
   		     ylabel={$a(t)$},
    	     xlabel={Time $(\si{s})$},
    	     grid=major,
             cycle list name=my markers
             ]
\addplot +[mark=none] table [x=t, y=a] {\optimal};
\addplot +[mark=none] table [x=t, y=a] {\nlopt};
\addplot +[only marks,thin] table [x=t, y=a] {\nloptcp};

\coordinate (bot) at (rel axis cs:1,0);% coordinate at bottom of the last plot
\end{groupplot}  % end
\end{tikzpicture}
\end{minipage}% Do not skip a line between minipages. Otherwise, it will place the figures one below the other.
\begin{minipage}[b]{0.5\linewidth}
\centering
\begin{tikzpicture}
\begin{groupplot}[group style={group name = myplot, group size = 1 by 2, horizontal sep = \hsp, vertical sep = \vsp}, height = \hb, width = \wb]
\nextgroupplot[
             xmin = 0,
             xmax = 1,
   		     ylabel={$\lambda_x(t)$},
   		     xticklabels=\empty,
    	     grid=major,
             cycle list name=my markers
             ]
\addplot +[mark=none] table [x=t, y=xcs] {\optimal}; \label{plots:bdd}
\addplot +[mark=none] table [x=t, y=xcs] {\nlopt}; \label{plots:bde}
\addplot +[only marks,thin] table [x=t, y=xcs] {\nloptcp}; \label{plots:bdf}

\coordinate (top) at (rel axis cs:0,1);% coordinate at top of the first plot

\nextgroupplot[
             xmin = 0,
             xmax = 1,
   		     ylabel={$\lambda_v(t)$},
   		     xlabel={Time $(\si{s})$},
    	     grid=major,
             cycle list name=my markers
             ]
\addplot +[mark=none] table [x=t, y=vcs] {\optimal};
\addplot +[mark=none] table [x=t, y=vcs] {\nlopt};
\addplot +[only marks,thin] table [x=t, y=vcs] {\nloptcp};

\coordinate (bot) at (rel axis cs:1,0);% coordinate at bottom of the last plot
\end{groupplot}  % end

% legend
\path (top|-current bounding box.north)--
      coordinate(legendpos)
      (bot|-current bounding box.north);
\matrix[
    matrix of nodes,
    anchor=south,
    draw,
    inner sep=0.2em,
    draw
  ]at([yshift=1ex]legendpos)
  {
    \ref{plots:bdd} & Optimal & [5pt] \\
    \ref{plots:bde} & \getdata[1]\NLstringsA & [5pt]\\
    \ref{plots:bdf} & \getdata[2]\NLstringsA & [5pt]\\};

\end{tikzpicture}
\end{minipage}
\if\slides0
	\caption{State, control, and costate trajectories using \NLOpt (with 30 LGR nodes) compared to the analytical optimal solution for the Bryson Denham problem \label{fig:bda}}
\fi
\end{figure}
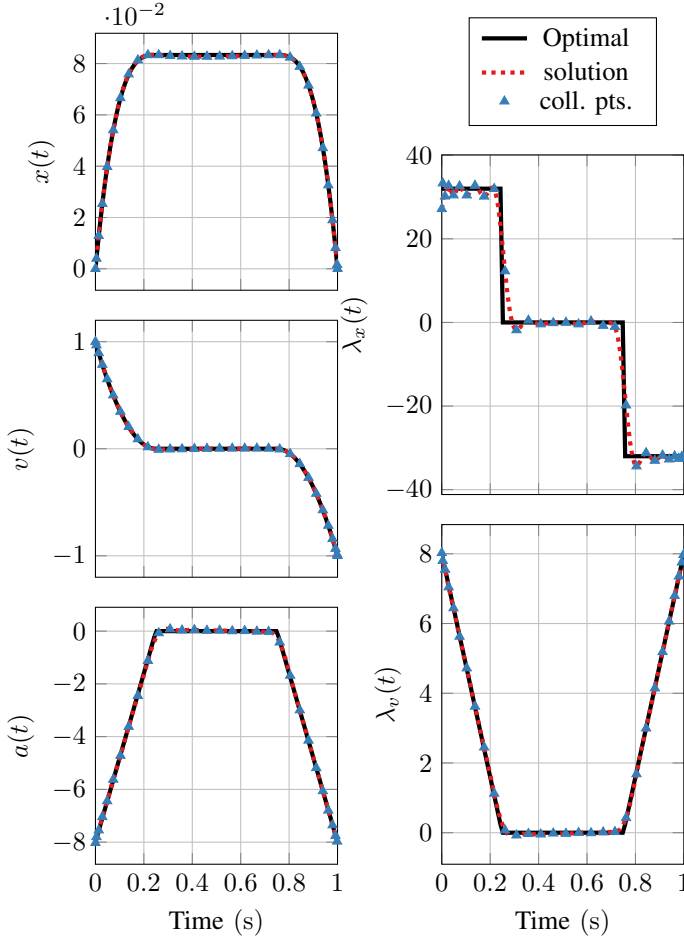
The NLP solver for this example is \IPOPT and an hp-method in \NLOpt is used with $4$ intervals and $10$ LGR nodes.
\section{Moon lander problem} \label{sec:moon}
\subsection{Closed-loop} \label{sec:moonCL}
Fig. \ref{fig:mlaa} shows the closed-loop solution to the moon lander problem using \NLOptnsp.
The closed-loop trajectory of the plant is very close  to the analytic solution.
Additionally, all of the solve-times are well below the chosen execution horizon $t_{ex}$ of $0.2\;\si{s}$; thus \NLOpt solves this NMPC problem in real-time.
% read in data and creatcat st_c	e macros
\pgfplotstableread{MoonLander/case1/results/sparse_optimal_solution.csv}{\optimal}

\pgfplotstableread{mpc/MoonLander/caseE/results/plant.csv}{\A}

\pgfplotstableread{mpc/MoonLander/caseE/results/opt.csv}{\opt}

%https://tex.stackexchange.com/questions/68577/compiling-a-document-with-pgfplots-processing-only-every-x-th-data-point

% https://tex.stackexchange.com/questions/272984/change-mark-size-in-legend-entry

\if\slides1
\newcommand{\ha}{4cm}
\newcommand{\wa}{6cm}

\newcommand{\hb}{5cm}
\newcommand{\wb}{6cm}
\else
\if\doublecol1
  \renewcommand{\ha}{5cm}
  \renewcommand{\wa}{4.8cm}

  \renewcommand{\hb}{6cm}
  \renewcommand{\wb}{4.8cm}
\else
  \renewcommand{\ha}{5cm}
  \renewcommand{\wa}{6cm}

  \renewcommand{\hb}{6.9cm}
  \renewcommand{\wb}{6cm}
\fi
\fi

\begin{figure}[!htb]
\begin{minipage}[b]{0.5\linewidth}
\centering
\begin{tikzpicture}

\begin{groupplot}[group style={group name = myplot, group size = 1 by 2, horizontal sep = 0.4cm, vertical sep = 1cm}, height = \ha, width = \wa]

\nextgroupplot[
   		     ylabel={solve-time $(\si{s})$},
             ylabel style={align=center, text width=2cm},
             xlabel={Iterations},
    	grid=major,
             cycle list name=my markers
             ]
\addplot +[mark=none] table [x=evalNum, y=tSolve] {\opt};

%\draw[dashed, very thick, blue] (0,0.2) -- (30,0.2);
%\node at (10,0.12) [fill=white] {Real-time threshold};

\nextgroupplot[
             xmin = 0,
             xmax = 4.2,
   		     ylabel={$x(t)$},
           xlabel={Time $(\si{s})$},
    	     grid=major,
             cycle list name=my markers solid
             ]
\addplot +[mark=none] table [x=t, y=x] {\optimal}; \label{plots:mpca}
\addplot +[mark=none] table [x=t, y=x] {\A}; \label{plots:mpcb}

\end{groupplot}  % end

\end{tikzpicture}
\end{minipage}% Do not skip a line between minipages. Otherwise, it will place the figures one below the other.
\begin{minipage}[b]{0.5\linewidth}
\centering
\begin{tikzpicture}
\coordinate (top) at (rel axis cs:0,1);% coordinate at top

\begin{groupplot}[group style={group name = myplot, group size = 1 by 2, horizontal sep = 0.4cm, vertical sep = 0.4cm}, height = \ha, width = \wa]

\nextgroupplot[
             xmin = 0,
             xmax = 4.2,
   		     ylabel={$v(t)$},
           xticklabels=\empty,
    	     grid=major,
             cycle list name=my markers solid
             ]
\addplot +[mark=none] table [x=t, y=v] {\optimal};
\addplot +[mark=none] table [x=t, y=v] {\A};

\nextgroupplot[
             xmin = 0,
             xmax = 4.2,
   		     ylabel={$a(t)$},
   		     xlabel={Time $(\si{s})$},
    	     grid=major,
             cycle list name=my markers solid
             ]
\addplot +[mark=none] table [x=t, y=a] {\optimal};
\addplot +[mark=none] table [x=t, y=a] {\A};

\end{groupplot}  % end

\coordinate (bot) at (rel axis cs:1,0);% coordinate at bottom of the last plot

% legend
\path (top|-current bounding box.north)--
      coordinate(legendpos)
      (bot|-current bounding box.north);
\matrix[
    matrix of nodes,
    anchor=south,
    draw,
    inner sep=0.2em,
    draw
  ]at([yshift=1ex,xshift=0ex]legendpos)
  {
    \ref{plots:mpca} & \getdata[1]\analytic   & [5pt] \\
    \ref{plots:mpcb} & \getdata[1]\MPCstringsB &[5pt] \\};
\end{tikzpicture}
\end{minipage}
\if\slides0
	\caption{Closed-loop trajectories for moon lander problem compared to the analytic solution \label{fig:mlaa}}
\fi
\end{figure}
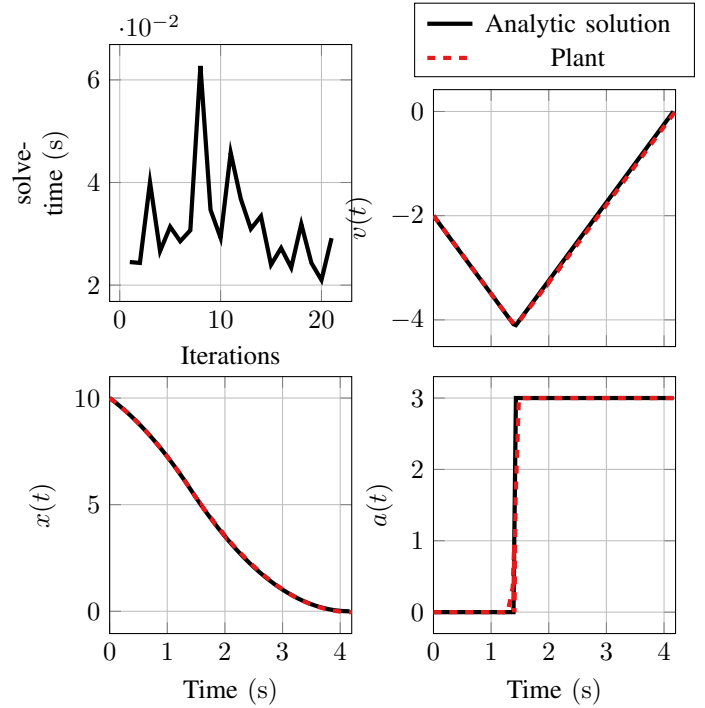
The NLP solver for this example is \IPOPT and an hp-method in \NLOpt is used with $4$ intervals and $10$ LGR nodes.

\subsection{Open-loop}
In Fig. \ref{fig:mla}, it can be seen that both \NLOpt and \PROPT determine the analytic solution accurately, with $30$ LGR and Chebyshev nodes, respectively.
However, there is an overshoot in the solution of the control with both \NLOpt and \PROPTnsp.
This is due to the bang-bang nature of the analytic solution.
It is noted that this overshoot may be mitigated using either mesh refinement \cite{patterson2015ph, darby2009mesh, jain2008trajectory} or radial basis functions \cite{hossein_2016}.
% read in data and creatcat st_c	e macros
\pgfplotstableread{MoonLander/case1/results/sparse_optimal_solution.csv}{\optimal}
\pgfplotstableread{MoonLander/case1/results/PROPT_30_poly_pts_filtered.csv}{\pr}
\pgfplotstableread{MoonLander/case1/results/PROPT_30_col_pts.csv}{\proptcp}
\pgfplotstableread{MoonLander/case1/results/st_ctr_poly.csv}{\nlopt}
\pgfplotstableread{MoonLander/case1/results/st_ctr.csv}{\nloptcp}

%https://tex.stackexchange.com/questions/68577/compiling-a-document-with-pgfplots-processing-only-every-x-th-data-point

% https://tex.stackexchange.com/questions/272984/change-mark-size-in-legend-entry

\if\slides1
\newcommand{\ha}{4cm}
\newcommand{\wa}{6cm}

\newcommand{\hb}{5cm}
\newcommand{\wb}{6cm}
\else
\if\doublecol1
  \renewcommand{\ha}{5cm}
  \renewcommand{\wa}{4.8cm}

  \renewcommand{\hb}{6cm}
  \renewcommand{\wb}{4.8cm}
\else
  \renewcommand{\ha}{5cm}
  \renewcommand{\wa}{6cm}

  \renewcommand{\hb}{6.9cm}
  \renewcommand{\wb}{6cm}
\fi
\fi

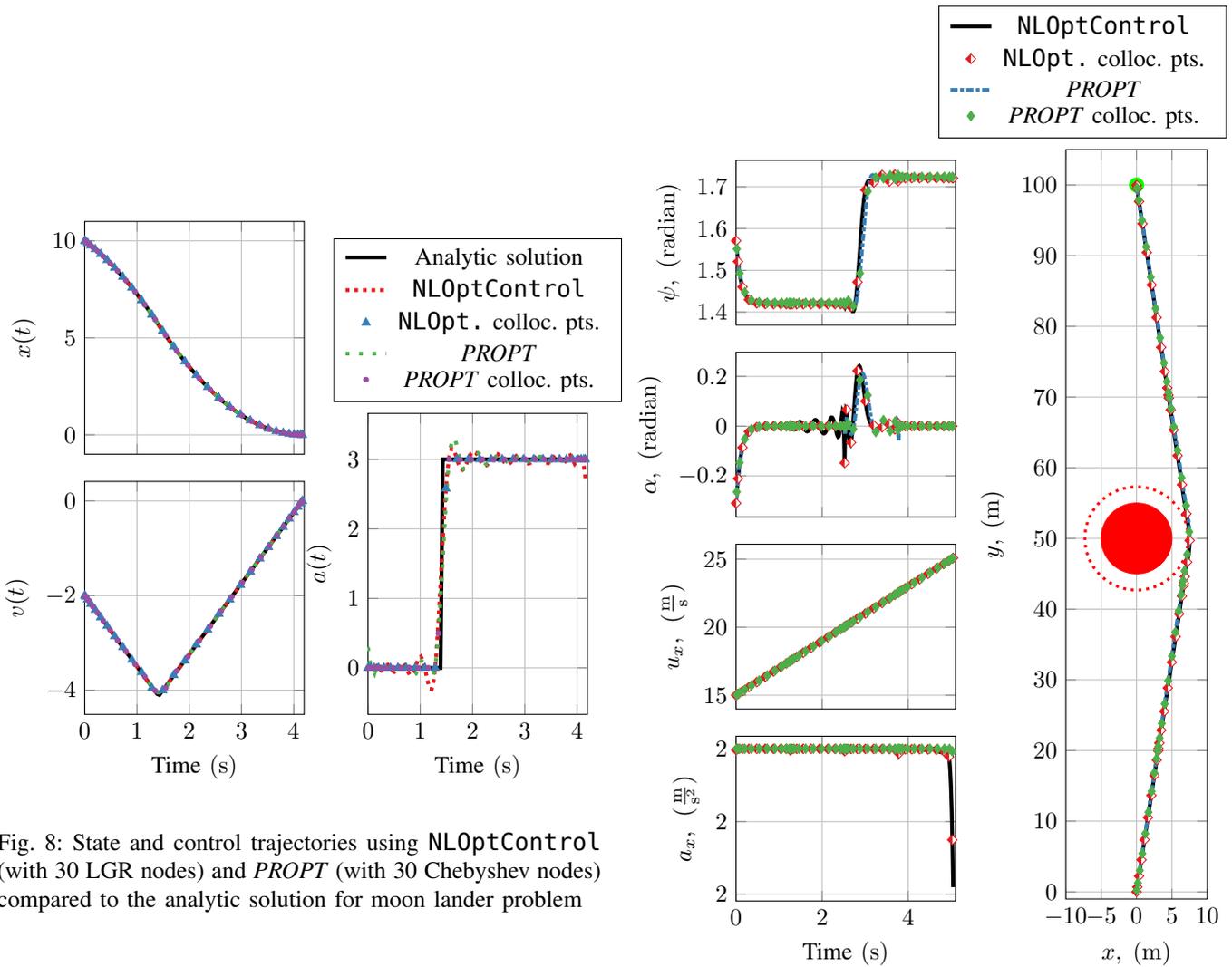
\begin{figure}[!htb]
\begin{minipage}[b]{0.5\linewidth}
\centering
\begin{tikzpicture}

\begin{groupplot}[group style={group name = myplot, group size = 1 by 2, horizontal sep = 0.4cm, vertical sep = 0.4cm}, height = \ha, width = \wa]

\nextgroupplot[
             xmin = 0,
             xmax = 4.2,
   		     ylabel={$x(t)$},
   		     xticklabels=\empty,
    	     grid=major,
             cycle list name=my markers
             ]
\addplot +[mark=none] table [x=t, y=x] {\optimal}; \label{plots:mla}
\addplot +[mark=none] table [x=t, y=x] {\nlopt}; \label{plots:mlb}
\addplot +[only marks,thin] table [x=t, y=x] {\nloptcp}; \label{plots:mlc}
\addplot +[mark=none] table [x=t, y=x] {\pr}; \label{plots:mld}
\addplot +[only marks,thin] table [x=t, y=x] {\proptcp}; \label{plots:mle}

\nextgroupplot[
             xmin = 0,
             xmax = 4.2,
   		     ylabel={$v(t)$},
   		     xlabel={Time $(\si{s})$},
    	     grid=major,
             cycle list name=my markers
             ]
\addplot +[mark=none] table [x=t, y=v] {\optimal};
\addplot +[mark=none] table [x=t, y=v] {\nlopt};
\addplot +[only marks,thin] table [x=t, y=v] {\nloptcp};
\addplot +[mark=none] table [x=t, y=v] {\pr};
\addplot +[only marks,thin] table [x=t, y=v] {\proptcp};
\end{groupplot}  % end

\end{tikzpicture}
\end{minipage}% Do not skip a line between minipages. Otherwise, it will place the figures one below the other.
\begin{minipage}[b]{0.5\linewidth}
\centering
\begin{tikzpicture}
\coordinate (top) at (rel axis cs:0,1);% coordinate at top
\begin{axis}[height = \hb,% 0.7\textheight,%7cm,
             width = \wb,%0.8\textwidth,%5cm,
             xmin = 0,
             xmax = 4.2,
   		     ylabel={$a(t)$},
   		     xlabel={Time $(\si{s})$},
    	     grid=major,
             cycle list name=my markers
             ]
\addplot +[mark=none] table [x=t, y=a] {\optimal};
\addplot +[mark=none] table [x=t, y=a] {\nlopt};
\addplot +[only marks,thin] table [x=t, y=a] {\nloptcp};
\addplot +[mark=none] table [x=t, y=a] {\pr};
\addplot +[only marks,thin] table [x=t, y=a] {\proptcp};
\end{axis}

\coordinate (bot) at (rel axis cs:1,0);% coordinate at bottom of the last plot

% legend
\path (top|-current bounding box.north)--
      coordinate(legendpos)
      (bot|-current bounding box.north);
\matrix[
    matrix of nodes,
    anchor=south,
    draw,
    inner sep=0.2em,
    draw
  ]at([yshift=1ex,xshift=0ex]legendpos)
  {
    \ref{plots:mla} & \getdata[1]\analytic   & [5pt] \\
    \ref{plots:mlb} & \getdata[1]\NLstringsC &[5pt] \\
    \ref{plots:mlc} & \getdata[2]\NLstringsC & [5pt] \\
    \ref{plots:mld} & \getdata[1]\PRstringsA & [5pt] \\
    \ref{plots:mle} & \getdata[2]\PRstringsA & [5pt] \\};
\end{tikzpicture}
\end{minipage}
\if\slides0
	\caption{State and control trajectories using \NLOpt (with 30 LGR nodes) and \PROPT (with 30 Chebyshev nodes) compared to the analytic solution for moon lander problem \label{fig:mla}}
\fi
\end{figure}

\section{Benchmark problem} \label{sec:obs}
This section provides an example of the type of solutions that are obtained from the benchmark between \NLOpt and \PROPTnsp.
For \NLOptnsp, the hp-method with LGR nodes and four intervals and $10$ collocation points per interval is used.
\PROPT is set to use four phases and $10$ collocation points per phase and Chebyshev nodes.
Fig. \ref{fig:bib} compares the results of these solvers, where it can be seen that position trajectories are close.
Starting at a speed of $15\;\si{\frac{m}{s}}$, the solutions obtained from both \PROPT and \NLOpt apply maximum acceleration from $t_0=0\;\si{s}$ to $t_f =\;5.1\;\si{s}$ while avoiding collision with the obstacle and reaching the desired goal position.
The trajectories for \NLOpt exhibit large oscillations in the $\alpha$ trajectory.
These oscillations may be an artifact of the Runge phenomenon and seem to be reduced with \PROPT as it uses Chebyshev nodes.

There is no analytic solution to this problem.

\pgfplotstableread{Bicycle/case1/results/PROPT_4_10_poly_pts.csv}{\pr}
\pgfplotstableread{Bicycle/case1/results/PROPT_4_10_col_pts.csv}{\proptcp}
\pgfplotstableread{Bicycle/case1/results/st_ctr_poly.csv}{\nlopt}
\pgfplotstableread{Bicycle/case1/results/st_ctr.csv}{\nloptcp}

\if\slides1
\newcommand{\ha}{3cm}
\newcommand{\wa}{7cm}

\newcommand{\hb}{5.5cm}
\newcommand{\wb}{6cm}

\else
  \if\doublecol1
    \renewcommand{\ha}{4cm}
    \renewcommand{\wa}{4.8cm}

    \renewcommand{\hb}{11cm}
    \renewcommand{\wb}{4.8cm}
  \else
    \renewcommand{\ha}{5.2cm}
    \renewcommand{\wa}{6.5cm}

    \renewcommand{\hb}{13.5cm}
    \renewcommand{\wb}{5cm}
  \fi
\fi

\begin{figure}[!htb]
% https://tex.stackexchange.com/questions/409919/creating-a-non-standard-array-of-plots/409936#409936
\begin{minipage}[b]{0.5\linewidth}
\centering
\begin{tikzpicture}

\begin{groupplot}[group style={group name = myplot, group size = 1 by 4, horizontal sep = 0.4cm, vertical sep = 0.4cm}, height = \ha, width = \wa]
\nextgroupplot[
             xmin = 0,
             xmax = 5.1,
   	ylabel={$\psi,\;(\si{radian})$},
   	ylabel style={align=center, text width=2cm},
    	grid=major,
    	xticklabels=\empty,
             cycle list name=my markers
             ]
\addplot +[mark=none] table [x=t, y=psi] {\nlopt}; \label{plots:bia}
\addplot +[only marks,thin] table [x=t, y=psi] {\nloptcp}; \label{plots:bib}
\addplot +[mark=none] table [x=t, y=psi] {\pr}; \label{plots:bic}
\addplot +[only marks,thin] table [x=t, y=psi] {\proptcp}; \label{plots:bid}

\nextgroupplot[
             xmin = 0,
             xmax = 5.1,
             ylabel={$\alpha,\;(\si{radian})$},
             ylabel style={align=center, text width=2cm},
    	grid=major,
    	xticklabels=\empty,
             cycle list name=my markers
             ]
\addplot +[mark=none] table [x=t, y=sa] {\nlopt};
\addplot +[only marks,thin] table [x=t, y=sa] {\nloptcp};
\addplot +[mark=none] table [x=t, y=sa] {\pr};
\addplot +[only marks,thin] table [x=t, y=sa] {\proptcp};

\nextgroupplot[
             xmin = 0,
             xmax = 5.1,
             ylabel={ $u_x,\;(\si{\frac{m}{s}})$},
             ylabel style={align=center, text width=2cm},
    	grid=major,
    	xticklabels=\empty,
             cycle list name=my markers
             ]
\addplot +[mark=none] table [x=t, y=ux] {\nlopt};
\addplot +[only marks,thin] table [x=t, y=ux] {\nloptcp};
\addplot +[mark=none] table [x=t, y=ux] {\pr};
\addplot +[only marks,thin] table [x=t, y=ux] {\proptcp};

\nextgroupplot[
             xmin = 0,
             xmax = 5.1,
   	ylabel={$a_x,\;(\si{\frac{m}{s^2}})$},
   	ylabel style={align=center, text width=2cm},
   	xlabel={Time $(\si{s})$},
    	grid=major,
             cycle list name=my markers
             ]
\addplot +[mark=none] table [x=t, y=ax] {\nlopt};
\addplot +[only marks,thin] table [x=t, y=ax] {\nloptcp};
\addplot +[mark=none] table [x=t, y=ax] {\pr};
\addplot +[only marks,thin] table [x=t, y=ax] {\proptcp};
\end{groupplot}  % end

\end{tikzpicture}
\end{minipage}% Do not skip a line between minipages. Otherwise, it will place the figures one below the other.
\begin{minipage}[b]{0.5\linewidth}
\centering
\begin{tikzpicture}
\coordinate (top) at (rel axis cs:0,1);% coordinate at top of the first plot
\begin{axis}[
	            height = \hb,
             xmin = -10,
             xmax = 10,
             ymin = -1,
             ymax = 105,
             scale only axis=true,
             axis equal image,
             ylabel={$y,\;(\si{m})$},
             xlabel={$x,\;(\si{m})$},
             grid=major,
             cycle list name=my markers,
             ]
\addplot +[mark=none] table [x=x, y=y] {\nlopt};
\addplot +[only marks,thin] table [x=x, y=y] {\nloptcp};
\addplot +[mark=none] table [x=x, y=y] {\pr};
\addplot +[only marks,thin] table [x=x, y=y] {\proptcp};

% obstacle
\draw[red, dotted, very thick] (axis cs:0,50) circle [radius=7.3];
\draw[red, fill] (axis cs:0,50) circle [radius=5];
%ellipse [x radius=1cm, y radius=1cm];

% \if\slides0
% 	\draw[<->,black,very thick] (0,43) to (0,45);
% 	\node [below] at (0,44) {$m$};
% 	\draw[<->,white,very thick] (0,50) to (-5,50);
% 	\node [above,white] at (-2.5,50) {$a_{obs}$};
% 	\draw[<->,white,very thick] (0,50) to (0,55);
% 	\node [right,white] at (0,52.5) {$b_{obs}$};
% \fi
% \draw[->,black,very thick] (-3,35) to (-2,47);
% \node [below] at (-3,35) {Obstacle};

% goal
\draw[green, fill] (axis cs:0,100) circle [radius=1];
\draw[->,black,very thick] (-6,95) to (0,99);
\node [below] at (-6,95) {Goal};

\end{axis}

\coordinate (bot) at (rel axis cs:1,0);% coordinate at bottom of the last plot

% legend
\path (top|-current bounding box.north)--
      coordinate(legendpos)
      (bot|-current bounding box.north);
\matrix[
    matrix of nodes,
    anchor=south,
    draw,
    inner sep=0.2em,
    draw
  ]at([yshift=1ex,xshift=-5ex]legendpos)
  {
    \ref{plots:bia} & \getdata[1]\NLstringsC &[5pt]\\
    \ref{plots:bib} & \getdata[2]\NLstringsC & [5pt] \\
    \ref{plots:bic} & \getdata[1]\PRstringsA & [5pt]\\
    \ref{plots:bid} & \getdata[2]\PRstringsA & [5pt]\\};
\end{tikzpicture}
\end{minipage}
\if\slides0
	\caption{State and control trajectories using \NLOpt (with 4 intervals and 10 LGR nodes) and \PROPT (with 4 intervals and 10 Chebyshev nodes) for the kinematic ground vehicle problem \label{fig:bib}}
\fi
\end{figure}

\if\thesis0
\end{appendices}
\fi

\end{document}